\documentclass[12pt]{article}
\usepackage{epsfig}
\usepackage{amsfonts}
\usepackage{amssymb}

\renewcommand{\thesection}{\arabic{section}}

\catcode`@=11
\@addtoreset{equation}{section}
\@addtoreset{equation}{subsection}
\def\theequation{\ifnum\value{section}=0 \arabic{equation}\ignorespaces
\else \ifnum\value{section}=-1 A.\arabic{equation}\ignorespaces
\else \ifnum\value{subsection}=0 \thesection.\arabic{equation}\ignorespaces
\else \thesection.\arabic{subsection}.\arabic{equation}\ignorespaces
                           \fi
                      \fi
                 \fi}
\catcode`@=12

\newcommand{\bq}{\begin{equation}}
\newcommand{\fq}{\end{equation}}
\newcommand{\bqr}{\begin{eqnarray}}
\newcommand{\fqr}{\end{eqnarray}}
\newcommand{\non}{\nonumber \\[1ex]}

\newcommand{\rf}[1]{(\ref{#1})}

\def\alp{\alpha}   \def\bet{\beta}    \def\gam{\gamma}
\def\del{\delta}   \def\eps{\epsilon} 
    \def\th{\theta}

  \def\cF{{\cal F}}

 \def\cN{{\cal N}} 
  
 \def\cT{{\cal T}}

\def\half{{1\over 2}}

\def\sfrac#1#2{{\textstyle{\frac#1#2}}}
\def\pit{{\textstyle{\pi\over\tau_2}}}
\def\ab{[{\textstyle{\alp\atop\beta}}]}
\def\hh{[{\textstyle{1/2\atop1/2}}]}
\def\oo{[{\textstyle{0\atop0}}]}
\def\dt#1{{\buildrel {\hbox{\LARGE .}} \over {#1}}}  
\def\ad{{\dt{\alpha}}}
\def\bd{{\dt{\beta}}}
\def\pd{{\dt{+}}}
\def\md{{\dt{-}}}

\newcommand{\C}{\mathbb C}
\newcommand{\R}{\mathbb R}
\newcommand{\Z}{\mathbb Z}

\begin{document}

\thispagestyle{empty}
\marginparwidth = .5in
% The following command moves parenthetical comments of the page.
\marginparsep = 1.2in

\begin{flushright}
\begin{tabular}{l}
ANL-HEP-PR-00-061 \\
ITP-UH-13/00 \\
hep-th/0007020
\end{tabular}
\end{flushright}

\vspace{12mm}
\begin{center}

{\Large\bf N=2 Quantum String Scattering}\\  

\vspace{14mm}

{\bf Gordon Chalmers}
\\[5mm]
{\em Argonne National Laboratory \\
High Energy Physics Division \\
9700 South Cass Avenue \\
Argonne, IL  60439-4815 } \\
{Email: chalmers@pcl9.hep.anl.gov}\\[5mm]  

\vskip .2in
{\bf Olaf Lechtenfeld} \ and \ {\bf Bernd Niemeyer} 
\\[5mm] 
{\em Institut f\"ur Theoretische Physik  \\
Universit\"at Hannover \\
Appelstra\ss{}e, 30167 Hannover, Germany }\\
{Email: lechtenf,niemeyer@itp.uni-hannover.de}

\vspace{15mm}

{\bf Abstract}  

\end{center}

We calculate the genus-one three- and four-point amplitudes in the 
$2{+}2$ dimensional closed $N{=}(2,2)$ worldsheet supersymmetric string 
within the RNS formulation.  Vertex operators are redefined with
the incorporation of spinor helicity techniques, and the quantum scattering
is shown to be manifestly gauge and Lorentz invariant after normalizing 
the string states.  The continuous spin structure summation over 
the monodromies of the worldsheet fermions is carried out explicitly, 
and the field-theory limit is extracted.  The amplitude in this limit 
is shown to be the maximally helicity violating amplitude in pure gravity 
evaluated in a two-dimensional setting, which vanishes, unlike the 
four-dimensional result.  The vanishing of the genus-one $N{=}2$ closed 
string amplitude is related to the absence of one-loop divergences in 
dimensionally regulated IIB supergravity.  Comparisons and contrasts between 
self-dual field theory and the $N{=}2$ string theory are made at the 
quantum level; they have different S-matrices.  Finally, we point to
further relations with self-dual field theory and two-dimensional models.  

\vfill

\setcounter{page}{0}
\newpage
\setcounter{footnote}{0}

\baselineskip=16pt

%%%%%%%%%%%%%%%%%%%%%%%%%%%%%%%%%%%%%%%%%%%%%%%%%%%%%%%%%%%%%%%%%%%%%%%%%

\section{Introduction}  

The field equations 
\bqr 
R_{\mu\nu}\ =\ {\tilde R}_{\mu\nu} \qquad\qquad\qquad 
F_{\mu\nu}\ =\ {\tilde F}_{\mu\nu}  
\label{selfduality} 
\fqr 
pertain to many aspects of physics and mathematical physics: 
self-dual field theory, string theory, instantons and monopoles, and 
the classification of four-manifolds.  

The $N{=}2$ worldsheet supersymmetric string is unique among string 
theories as its critical dimension is four \cite{Ademollo:1976an, 
Ademollo:1976pp, Bruce:1976ng, Fradkin:1981dd, D'Adda:1987rx}.  Its 
full spectrum and its exact (in $\alp'$) classical field equations
have been identified to be merely those of self-dual gravity and 
self-dual Yang-Mills theory \cite{Ooguri:1990ww, Ooguri:1991fp}
(for a review up to 1992, see \cite{Marcus}).  
Several quantum field theory formulations of the latter theories point 
to a non-vanishing S-matrix at the quantum level \cite{Chalmers:1996rq}, 
a fact which is, however, in contrast to the claims of zero quantum 
S-matrix for the $N{=}2$ string 
\cite{Berkovits:1995vy, Berkovits:1995ym, Junemann:2000xj}.
Apparently, we witness a quantum discrepancy between two theories which 
are classically equivalent.  In this work we address this question by 
calculating the $N{=}(2,2)$ closed-string genus-one amplitudes in the 
RNS formulation and identifying the target spacetime theory which gives 
rise to these amplitudes.  

Self-duality in $d=2{+}2$ dimensions (or in $d=4{+}0$) is implemented in 
the field equations of gravity and Yang-Mills by \rf{selfduality}, of 
which the only known Lorentz covariant Lagrangian formulation employs
Lagrange multipliers and is given by~\footnote{
Sub- and superscripts $\alp\in\{+,-\}$ and $\ad\in\{\pd,\md\}$ are
spinor indices of $SL(2,\R)$ (or $SU(2)$).}
\bqr  
{\cal L}\ =\ {\rm Tr} ~G^{\alpha\beta} F_{\alpha\beta} 
\qquad\qquad\qquad 
{\cal L}\ =\ {\rm Tr} ~\rho^{\alpha\beta} R_{\alpha\beta} \quad , 
\label{lcsdym}
\fqr  
where $F$ and $R$ are the self-dual projections of the field-strength
and Riemann tensor for a gauge and spin connection vector, respectively.
\cite{Siegel:1993wd, Chalmers:1996rq}. 
The Lagrangian \rf{lcsdym} involves two fields, related to the two 
polarization states of a gauge field, yet only one appears as
an asymptotic state.

Alternatively, fixing a light-cone gauge in \rf{selfduality} allowed
Leznov \cite{Leznov:1988up} and Plebanski \cite{Plebanski} to reduce the 
self-duality equations to a pair of second-order equations
\bqr
-\square\phi + \sfrac{g}{2}[\partial_+{}^\ad \phi\,,\,\partial_{+\ad}\phi] =0
\qquad\qquad
-\square\psi + \sfrac{\kappa}{2}\,\partial_+{}^\ad \partial_+{}^\bd \psi\;
\partial_{+\ad} \partial_{+\bd} \psi =0
\label{leznovframe}
\fqr
for scalar prepotentials $\phi$ (to $F$) and $\psi$ (to $R$) which extremize 
the respective Lorentz non-covariant gauge-fixed actions belonging to
\bqr 
{\cal L}\ =\ {\rm Tr}~ \phi\,\Bigl( -\sfrac12 \square\phi + \sfrac{g}{6} 
[ \partial_+{}^\ad \phi\,,\,\partial_{+\ad} \phi ] \Bigr)
\label{Leznovform}
\fqr 
and 
\bqr  
{\cal L}\ =\ \psi\,\Bigl( -\sfrac12 \square \psi + \sfrac{\kappa}{6}\, 
\partial_+{}^\ad \partial_+{}^\bd \psi\;\partial_{+\ad} \partial_{+\bd}\psi
\Bigr) \ .
\label{secondplebanski} 
\fqr 
Further Lorentz non-covariant formulations of self-dual quantum field theories
can be found by solving the gauge constraints in \rf{selfduality} differently
\cite{Yang,Nair:1990aa}.  As these one-field actions share a coupling constant
of positive length dimension they are all power-counting non-renormalizable.  

The Lorentz-covariant two-field actions~\cite{Chalmers:1996rq} are much better
behaved in this respect. In light-cone gauge, their Lagrangians are
\bqr
{\cal L}\ =\ {\rm Tr}~ \tilde\phi\,\Bigl( - \square\phi + \sfrac{g}{2}
[ \partial_+{}^\ad \phi\,,\,\partial_{+\ad} \phi ] \Bigr)
\label{twofield1}
\fqr
and
\bqr
{\cal L}\ =\ \tilde\psi\,\Bigl( - \square \psi + \sfrac{\kappa}{2}\,
\partial_+{}^\ad \partial_+{}^\bd \psi\;\partial_{+\ad} \partial_{+\bd}\psi
\Bigr) 
\label{twofield2}
\fqr
which allows no scattering beyond one-loop, because the multiplier fields
go with $1/\hbar$.  The one- and two-field theories both
generate the maximally helicity violating (MHV) scattering at one-loop and 
the vanishing next-to-MHV amplitudes at tree-level \cite{Chalmers:1996rq}, 
and the latter theories are one-loop exact perturbatively. 

To compare with, the $N{=}2$ superstring has been shown 
(modulo contact term ambiguities) to possess trivial scattering in its 
critical dimension \cite{Berkovits:1995ym}.  This indicates the presence 
of an anomaly in the string, or a target-space interpretation different from 
self-dual gravity or gauge theory.  A possible anomaly interpretation behind 
the $d=3{+}1$ MHV amplitudes in gauge theory was initially pointed out in 
\cite{Bardeen:1996gk} in the context of the conserved symmetries of the field 
equations.  

Until now, the $N{=}2$ string quantum amplitude has never been computed 
in the traditional RNS formalism (but functional methods for the quantization 
at higher genera have been developed \cite{Bischoff:1997bn}).  However, 
by embedding the $N{=}2$ string in an $N{=}4$ topological string 
it was demonstrated that, up to contact terms, 
these amplitudes vanish to all loops \cite{Berkovits:1995vy, Berkovits:1995ym}.
Linearized symmetry arguments have also formally shown this in the RNS 
formulation in \cite{Junemann:2000xj}.  In  order to compare with the 
field-theory results and to find the root of this discrepancy, 
an explicit traditional computation at genus one is worthwhile. 
In the present work we perform this calculation.  We find that 
the $N{=}2$ string loop dynamics appears to be reduced to two dimensions.  
Based on earlier one-loop computations of the partition function~\cite{MM}
and the three-point function~\cite{Bonini:1991bf}, Marcus~\cite{Marcus}
already identified the technical origin of this dimensional mismatch.
Here, we confirm his observation and extend it to the full quantum dynamics 
by evaluating the one-loop four-point scattering.
As the $N{=}2$ string has a critical dimension of four,
with four (real) target spacetime coordinates, this calculation indicates 
that it represents a ghost system in the MHV sector of gauge theory.  In 
order to make the above explicit we render the scattering 
manifestly Lorentz invariant by normalizing the vertex operators and 
incorporating the gauge invariance through the use of spinor helicity 
techniques.  

A further unexpected relation arises between the one-loop MHV amplitude 
in pure gravity regulated to two dimensions and the next-to-MHV amplitude 
in IIB supergravity evaluated in ten dimensions.\footnote{
This dimension-shifting relation involving a change in the number of 
supersymmetries was initially found in \cite{Bern:1997ja}.}  
The absence of one-loop divergences in the massless sector of IIB supergravity 
in ten dimensions within dimensional reduction explains the vanishing of the 
two-dimensional MHV result and thereby the triviality of the field theory 
limit of the $N{=}2$ string scattering.  Alternatively, a relation
is found between two string theories: the IIB superstring in ten dimensions 
and the $N{=}2$ closed string in four dimensions.
Such a relation may originate in an integrable structure 
in the ultra-violet regime for the massless modes of the string 
(a supergravity analog of the Regge kinematical limit of Yang-Mills theory). 
If this connection extends to multi-loops, the vanishing theorems of the
$N{=}2$ string at higher genera deserve further study.

The outline of this work is as follows.  In section~2 we review and 
discuss the properties of the gauge theory MHV amplitudes in different 
dimensions.  Section~3 implements gauge invariance directly into the 
$N{=}2$ string scattering through the incorporation of spinor helicity 
techniques and a normalization of the vertex operators.  We also analyze 
contact term subtleties in the scattering.  In section~4 the three-point 
genus-one closed-string amplitude is obtained and compared with the field 
theoretic one.  In section~5 we finally compute a modular-integral expression 
for the genus-one closed-string four-point amplitude, carefully taking into 
account the superconformal ghost structure.  From this result, we extract
the field theory limit by taking $\alpha'\to0$ before summing over spin
structures (which then trivializes). The answer is zero, and the comparison
with the MHV amplitudes is made.  In section~6 we explicitly perform the 
spin structure summation on the full modular integrand before taking the 
field theory limit, with identical (vanishing) result.
A discussion and an Appendix on Jacobi theta functions conclude the paper.

%%%%%%%%%%%%%%%%%%%%%%%%%%%%%%%%%%%%%%%%%%%%%%%%%%%%%%%%%%%%%%%%%%%%%%%%%%

\section{Review of MHV Gauge Theory Amplitudes} 

Recent developments in techniques in gauge theory calculations~\footnote{
For a review at tree-level see \cite{mpreview} 
and at loop-level \cite{bdkreview}.} 
have made possible the calculation of closed analytic forms of several infinite 
sequences of one-loop gauge theory amplitudes. The maximally helicity violating 
(MHV) amplitudes are described by scattering of gauge fields of identical 
helicity, either in Yang-Mills theory or in gravity.  One of the 
features of these amplitudes is that in a supersymmetric theory they 
are identically zero to infinite loop order; 
this implies that at tree-level the amplitudes are identically zero.  
The amplitudes closest  to MHV are simpler to calculate, 
and the self-dual description has lead to 
reformulations and improved diagrammatic techniques in calculating gauge 
theory amplitudes \cite{Chalmers:1999jb} as well as second-order formulations 
for incorporating fermions~\footnote{
In a self-dual non-abelian background, 
fermions may be bosonized, and fields become spin independent.} 
\cite{Morgan:1995te,Chalmers:1999ui}.  
In this section we briefly review these maximally helicity violating 
amplitudes and describe their relations to both self-dual field theory and 
string theory.  The continuation of the four-point MHV gravity amplitude
and its conjectured form to $n$-point order to arbitrary dimensions 
is directly related to the zero-slope limit of the $N{=}2$ closed string.   

At one-loop in four dimensions, the leading-in-color partial amplitude 
for the scattering of $n$ gluons of identical out-going helicity in 
Yang-Mills theory is \cite{Bern:1994qk,Mahlon:1994si}
\bqr 
A_{n;1}^{[1]}(k_i)\ =\ 
-{i\over 48\pi^2} \sum_{1\leq i<j<k<l\leq n} {\langle ij\rangle 
[ jk] \langle kl\rangle [ li] \over 
\langle 12\rangle \langle 23\rangle \cdots \langle n1\rangle} 
\label{allplusgauge}
\fqr  
where the superscript $[J]$ represents the spin of the internal state 
(gluon, complex scalar or Weyl fermion give the same result up to a 
minus sign for half-integral spin).   The amplitude is written in 
color-ordered form \cite{colordecomp}; the leading-in-color group theory 
structure, 
\bqr  
N^2 ~ {\rm Tr}~ {\rm T}^{a_1} {\rm T}^{a_2} \ldots {\rm T}^{a_n} \ , 
\label{colorstruct} 
\fqr 
has been extracted from the kinematics in accord with Chan-Paton assignments 
in open string theory and gauge theory.  
In \rf{allplusgauge} we have decomposed each lightlike momentum vector $k_i$ 
into two momentum Weyl spinors and defined two different inner products,
\bqr 
k_i^{\alp\ad}= k_i^\alp k_i^\ad \qquad {\rm and} \qquad
\langle ij\rangle = k_i^\alpha k_{j,\alpha} \qquad 
[ij] = k_i^{\dot\alpha} k_{j,\dot\alpha} \ .
\label{spinorproducts}
\fqr 
In $2{+}2$ dimensions, 
$\langle ij\rangle$ is not the complex conjugate of $[ij]$; 
the Lorentz group is $SL(2,\R)\times SL(2,\R)'$ as opposed to $SL(2,\C)$ 
in $3{+}1$ dimensions (and there are poles in \rf{allplusgauge} in the 
self-dual plane parameterized by the $SL(2,\R)$ half of the Lorentz group).  

The analogous result for the all-plus gravitational amplitude 
\cite{Grisaru:1980re, Dunbar:1995bn}, 
%\bqr  
%A_{4}^{[2]}(k_i)\ =\ 
%-{\kappa^4\over 48\pi^2} {[12] [23] [34] [41] \over 
%\langle 12\rangle \langle 23\rangle \langle 34\rangle \langle 41\rangle }  
%\langle 12\rangle [23]\langle 34\rangle [41] 
%\ +\ {\rm perms} \ ,
%\label{fourpointgrav} 
%\fqr 
\bqr  
A_{4}^{[2]}(k_i)\ =\ 
-i \Bigl({\kappa\over 2}\Bigr)^4 {1\over 120\,(4\pi)^2}\, \Bigl(
{s_{12} \, s_{23} \over 
\langle12\rangle \langle23\rangle \langle34\rangle \langle41\rangle }
\Bigr)^2 \, (s_{12}^2 + s_{23}^2 + s_{13}^2) \ ,
\label{fourpointgrav} 
\fqr
and its $n$-point form \cite{Bern:1998xc}, 
together with the three-point vertex,
describe the scattering of all-plus helicity gravitons to one-loop order.  
The amplitude in \rf{allplusgauge} and its gravitational analog have a
number of features in common with $N{=}2$ string scattering. They are
channel-dual in the sense that exchange of any two legs gives the same form.
Furthermore, they have only two-particle poles (in one $SL(2,\R)$ factor
of the Lorentz group), which signals integrable characteristics 
related to the infinite number of symmetries in the self-dual field equations.

The $n$-point gauge theory amplitude in \rf{allplusgauge} has been 
found by constraining the functional form based on analyticity 
\cite{Bern:1994qk} as well as through a direct calculation with a fermion 
in the loop \cite{Mahlon:1994si}.  The amplitude in \rf{allplusgauge} also 
arises in a one-loop S-matrix element for self-dual Yang-Mills theory.
This happens for the Lorentz-covariant two-field theory (one-loop exact)
\cite{Chalmers:1996rq} as well as, to a factor of two, 
for the one-field (Leznov) formulation \cite{Chalmers:1996rq,Cangemi:1997rx}, 
although the latter is not Lorentz covariant (or one-loop exact). 
The same story occurs in gravity \cite{Bern:1998xc} where 
\rf{fourpointgrav} and its generalizations describe quantum self-dual 
gravity at one-loop \cite{Chalmers:1996rq}.  One might expect to find 
similar non-vanishing scattering amplitudes at the one-loop level in 
the $N{=}2$ string, as both the string and the self-dual field theory 
share the same classical field equations.  However, this expectation is 
not borne out by our calculation below.   

The $d$ dimensional generalization of the Yang-Mills result in 
\rf{allplusgauge} has been found in \cite{Bern:1997ja} up to six-point 
(together with a conjectured form at $n\geq 7$ point), 
and we list here the form of these amplitudes. At four-point one has
\bqr 
A_{4;1}^{[1]}(k_i)\ =\ {-2i\over \langle 12\rangle \langle 23\rangle 
\langle 34\rangle \langle 41\rangle} {(4-d)(2-d)\over 4(4\pi)^2}\, 
s_{12} s_{23} ~ I_4^{4+d}(s,t) \ , 
\label{ddimfourpoint}
\fqr 
where the box diagram $I_4^{4+d}$ is the integral function 
\bqr 
I_4^{p}(s,t)\ =\ \int\!{d^p\ell\over (2\pi)^p}\; 
{1\over \ell^2 (\ell-k_1)^2 (\ell-k_1-k_2)^2 (\ell+k_4)^2} \ ,
\fqr 
continued from $p$ to $d{+}4$ dimensions but with the 
external vectors in $d$ dimensions.  The generalization of 
the series in \rf{allplusgauge} arises by keeping the external 
kinematics and polarizations in four dimensions and analytically continuing 
the scalar integral functions. In a Schwinger proper-time formulation of the 
integrals this amounts to inserting additional factors of $\tau_2$ in the 
integral over the proper time.  In {\it four\/} dimensions, the $8$-dimensional 
box diagram in \rf{ddimfourpoint} relevant to the amplitude is UV divergent, 
but the result is finite because the $d{-}4$ prefactor extracts 
the residue.  In {\it two\/} dimensions the $6$-dimensional box 
diagram with external massless kinematics is both IR and UV finite, but the 
prefactor forces the result in \rf{ddimfourpoint} to be identically zero.  
The MHV amplitude thus vanishes upon continuation to $d{=}2$, 
without recourse to spacetime supersymmetry.  

The five- and six-point amplitudes and their dimensional form have the 
same properties as the expression \rf{ddimfourpoint}, as does the conjectured 
$n$-point form at one-loop.  The five-point amplitude,  
\bqr  
A_{5;1}^{[1]}(k_i) &=& {-i\over \langle 12\rangle \langle 23\rangle 
\langle 34\rangle \langle 45\rangle \langle 51\rangle} 
{(4-d)(2-d)\over 4(4\pi)^{d/2}} \Bigl[ s_{23} s_{34} I_4^{d+4} 
+ s_{34} s_{45} I_4^{d+4}  
\non
&+& \!\!\! s_{45} s_{51} I_4^{d+4} + 
s_{51} s_{12} I_4^{d+4} + s_{12} s_{23} I_4^{d+4} + 
4 i d \epsilon_{\mu\nu\rho\sigma} k_1^{\mu} k_2^{\nu} 
k_3^{\rho} k_4^{\sigma} I_5^{d+6} \Bigr] \,, 
\label{fivepointdim}
\fqr 
is zero when continued to two dimensions because the six-dimensional 
box and eight-dimensional pentagon in \rf{fivepointdim} are finite 
and the pre-factor vanishes in $d{=}2$.  The gauge theory result at 
six-point is similar and is described in eqs. (16) and (17) of reference 
\cite{Bern:1997ja}.  

In (spacetime) {\it supersymmetric\/} gauge or gravitational theory, the 
MHV one-loop amplitudes vanish because of a cancellation between the 
contributions stemming from different spin states running inside the loop 
\cite{superidentity}. Concretely,
\bqr
A^{[1]}\ =\ A^{[0]}\ =\ - A^{[{1\over 2}]}\ =\ A^{[2]} \ ,
\label{spectralflow}
\fqr 
for a gauge boson, complex scalar, Weyl fermion, or graviton,
so that amplitudes need to be computed for only one conveniently chosen
spin value.  

The all-$n$ conjectured form of the MHV Yang-Mills amplitude 
relates to a $d{+}4$ $\cN{=}16$ supersymmetric non-MHV amplitude 
as follows, 
\bqr  
A^{[0]}_{n;1}(k_i)\Big\vert_d = {(4-d)(2-d)\over 2}(4\pi)^2  
 {1\over \langle 12\rangle^4}
A_{n;1}^{\cN=16}(k_1^-,k_2^-,k_3^+,\ldots ,k_n^+) \Big\vert_{d+4} \ ,
\label{allnpoint}
\fqr 
where for definiteness we denote it for an internal complex scalar, 
with $[J{=}0]$.  The factor of $\langle 12\rangle^4$ gives the left-hand 
side the appropriate spinor weight to describe the negative-helicity gluons 
on legs one and two.  Curiously, the prefactor in \rf{allnpoint} is negative 
for $2\leq d\leq 4$.  Again, the finiteness of the amplitude on the right-hand 
side of \rf{allnpoint} in $d{+}4{=}6$ translates into the vanishing of the MHV 
amplitude in $d{=}2$.  For $d{+}4{=}8$ the UV singularity of $A_n^{\cN=16}$ 
reproduces \rf{allplusgauge}.

The explicit result in $d$ dimensions for the four-point one-loop maximally 
helicity violating Einstein-Hilbert gravitational amplitude is 
\bqr  
A_{4}^{[2]} (k_i) \Big\vert_d = {(4-d)(2-d)\,d\,(2+d) \over 8} (4\pi)^4 
{1\over \langle 12 \rangle^8}
A_{4}^{\cN=32}(1^{--},2^{--},3^{++},4^{++}) \Big\vert_{d+8} 
\label{gravfourrelation}
\fqr  
where the relation is between an MHV amplitude in $d$ dimensions 
to a non-MHV amplitude in $d{+}8$ dimensions and in the $\cN{=}32$ 
(maximally) supersymmetric theory.  Similar to the Yang-Mills 
case, the additional $\langle 12\rangle^4$ gives the MHV amplitude the 
proper helicity weight (the graviton has twice the spin) and dimensions.  

For $d{=}2$ the amplitude on the right-hand side in \rf{gravfourrelation} 
is to be evaluated in ten dimensions.  In this case
no counterterms occur in the amplitude calculation in dimensional regulation 
since the divergences at four-point are proportional to 
\bqr  
\left( {1\over d-10}\right) (s+t) + 
\left( {1\over d-10}\right) (t+u) + 
\left( {1\over d-10}\right) (u+s) 
\fqr 
which is zero on-shell, forcing the MHV result in \rf{gravfourrelation} in 
$d{=}2$ to vanish. Parallel to the relation in \rf{allnpoint} and 
generalizing \rf{gravfourrelation}, the conjectured $d$-dimensional 
gravitational MHV amplitude at arbitrary $n$-point coincides with the
$\cN{=}32$, $d{+}8$ next-to-MHV amplitude. The absence of a counterterm at 
$n$-point in the dimensionally regularized/reduced form of IIB supergravity
in ten dimensions means that, due to the prefactor in the $n$-point 
generalization of \rf{gravfourrelation}, the MHV result for graviton 
scattering in two dimensions is zero at arbitrary $n$-point order at one-loop. 

Two-dimensional gravity and Yang-Mills theory are topological and have no 
dynamical degrees of freedom.  The scattering in these theories is trivial in 
topologically trivial spacetime, which explains the vanishing of the 
amplitudes not only at one-loop but also to infinite loop order.  
A possible relation between the reduced form of the scattering in $d{=}2$ 
and that in $d{=}10$ implies further non-trivial structure in the 
ultra-violet of IIB supergravity.  

In the following we shall relate the above $d{=}2$ result
in the gravitational case to the scattering obtained in the RNS
formulation of the closed $N{=}2$ superstring in the zero-slope limit.
Given the holomorphic/anti-holomorphic factorization of the string integrand,
this relation might persist to the open string as well.

%%%%%%%%%%%%%%%%%%%%%%%%%%%%%%%%%%%%%%%%%%%%%%%%%%%%%%%%%%%%%%%%%%%%%%%%

\section{N=2 String Vertex Operators}  

In this section we review the relevant facts of the closed $N{=}2$ string
and its tree-level scattering amplitudes.
We pay particular attention to the its vertex operators, for two reasons:
First, the representation of the vertex operators affects possible contact
interactions and their contributions to scattering amplitudes.
Second, the normalization of the vertex operators translates to the
choice of external leg factors which are crucial to achieve
a manifestly gauge-invariant representation of the amplitudes via
spinor helicity techniques.  For a brief review, the reader may consult 
\cite{dubna} and references therein. 

%%%%%%%%%%%%%%%%%%%%%%%%%%%%
\bigskip

\noindent
{\bf 3.1\ \ Generalities}

\smallskip

{}From the worldsheet point of view, critical closed $N{=}2$ strings
in flat Kleinian space $\R^{2,2}$
are a theory of $N{=}(2,2)$ supergravity on a $1{+}1$ dimensional
(pseudo) Riemann surface, 
coupled to two chiral $N{=}(2,2)$ massless matter multiplets $X^a$, $a=1,2$.
The latter's components are complex scalars $x$ (the four string coordinates),
$SO(1,1)$ Dirac spinors $\psi$ (their four NSR partners) 
and complex auxiliaries $F$,
\bqr
X^a\ =\ 
x^a +\theta^- \psi^{+a} +\theta^+ \psi^{-\bar{a}} +\theta^+\theta^- F^a 
\fqr
with arguments $y\equiv z+\theta^+\theta^-$.
Complex conjugation reads
\bqr
z^*=z \qquad\qquad (\theta^+)^* = \theta^- \qquad\qquad
(x^a)^* = x^{\bar{a}} \qquad\qquad (\psi^{+a})^* = \psi^{-\bar{a}}
\label{cc}
\fqr
while chiral conjugation exchanges right- and left-movers via
\bqr
z \to \bar{z} \qquad\qquad \theta^\pm \to \bar{\theta}^\pm \qquad\qquad
x^a \to x^a \qquad\qquad \psi^{+a} \to \bar{\psi}^{+a} \ .
\fqr
The extended worldsheet supersymmetry has induced a spacetime complex structure
which reduces the global Lorentz symmetry, 
\bqr
{\rm Spin}(2,2)\ =\ SU(1,1) \times SU(1,1)' \
\longrightarrow\ U(1) \times SU(1,1)' \ \simeq\ U(1,1) \ .
\label{break1}
\fqr
In superconformal gauge, however, manifest $SO(2,2)$ symmetry is restored
in the worldsheet action, which is given by
\bqr  
S= \int\! d^2z\, d^2\theta d^2{\bar\theta} ~ K(X,{\bar X}) 
= \int\! d^2z\, \eta_{a\bar{a}}\,[
\partial x^a \bar\partial x^{\bar a} +
\psi^{+a} \bar\partial \psi^{-\bar{a}} +
\bar{\psi}^{+a} \partial \bar{\psi}^{-\bar{a}} ]
\label{gfaction}
\fqr
where $\eta_{a\bar{a}}={\rm diag}(+-)$ is the flat metric in~$\C^{1,1}$,
and the auxiliary fields have been integrated out.  

Although the above notation makes transparent the local R symmetry properties 
of the fields (for instance, $x$ is neutral while $\psi^\pm$ is not),
it is not convenient for our computations.
The interrelation \rf{cc} with complex conjugation allows us to change it,
\bqr
x^a \to x^{+a} \qquad\qquad x^{\bar{a}} \to x^{-a} \qquad\qquad
\psi^{+a} \to \psi^{+a} \qquad\qquad \psi^{-\bar{a}} \to \psi^{-a} \ ,
\fqr
so that the $SO(2,2)$ invariant scalar product reads
\bqr
k \cdot x\ =\ \sfrac12\, (k^+ \cdot x^- + k^- \cdot x^+)
\ =\ \sfrac12\, (k^{+1}x^{-1} - k^{+2}x^{-2} + k^{-1}x^{+1} - k^{-2}x^{+2}) 
\fqr
where the dot is also used to denote the $SU(1,1)'$ invariant scalar product.
There exist three {\it antisymmetric\/} $SU(1,1)'$ invariant products,
\bqr
k^+\wedge x^+ &=& \eps_{ab}\; k^{+a}x^{+b} = k^{+1}x^{+2} - k^{+2}x^{+1}
\non
k^+\wedge x^- &=& \sfrac12\, (k^+ \cdot x^- - k^- \cdot x^+)\
=\ \sfrac12\, (k^{+1}x^{-1} - k^{+2}x^{-2} - k^{-1}x^{+1} + k^{-2}x^{+2})
\non
k^-\wedge x^- &=& \eps_{ab}\; k^{-a}x^{-b} = k^{-1}x^{-2} - k^{-2}x^{-1}
\fqr
which feature prominently in the following.

The $N{=}(2,2)$ supergravity multiplet defines a gravitini and a Maxwell
bundle over the worldsheet Riemann surface.  The topology of the total space 
is labeled by the Euler number~$\chi$ of the punctured Riemann surface and
the first Chern number (instanton number)~$M$ of the Maxwell bundle.
It is notationally convenient to replace the Euler number by the ``spin''
\bqr
J\ :=\ -2\chi\ =\ 2n-4+4(\# {\rm handles})\ \in\ 2\Z \quad.
\fqr
The action \rf{gfaction} is to be considered for string worldsheets 
of a given topology.\footnote{
Of course, the Lagrangian in \rf{gfaction} is 
in general not correct globally.}  
The first-quantized string path integral for the $n$-point function $A_n$ 
includes a sum over worldsheet topologies~$(J,M)$,
weighted with appropriate powers in the string couplings~$(\kappa,e^{i\th})$:
\bqr
A_n(\kappa,\th)\ =\ \sum_{J=2n-4}^{\infty} \kappa^{J/2}\,A_n^J(\th)\ =\
\sum_{J=2n-4}^{\infty}\sum_{M=-J}^{+J}\kappa^{J/2}\,e^{iM\th}\,A_n^{J,M}
\fqr
where the instanton sum has a finite range because bundles with $|M|{>}J$
do not contribute.  The presence of Maxwell instantons breaks the explicit 
$U(1)$ factor in \rf{break1} but the $SU(1,1)$ factor (and thus the whole 
${\rm Spin}(2,2)$) is fully restored if we let
$\kappa^{1/4}(e^{i\th/2},e^{-i\th/2})$ transform as an $SU(1,1)$ spinor.
As a consequence, the string couplings depend on the $SO(2,2)$ Lorentz frame,
and we may choose a convenient one for calculations. 
We call the choice $\th{=}0$ a `Leznov frame' 
and name an averaging over~$\th$ a `Yang frame'.
The partial amplitudes $A_n^{J,M}$ are integrals over the metric,
gravitini, and Maxwell moduli spaces. The integrands may be obtained
as correlation functions of vertex operators in the $N{=}(2,2)$ superconformal
field theory on the worldsheet surface of fixed shape (moduli) and
topology.

The vertex operators produce from the (first-quantized) vacuum state the 
asymptotic string states in the scattering amplitude under consideration.
They correspond to the physical states of the $N{=}2$ closed string
and carry their quantum numbers.  Being representatives of the (semi-chiral) 
BRST cohomology, they are unique only up to BRST-trivial terms and 
normalization.  The physical subspace of the $N{=}2$ string Fock space in a 
covariant quantization scheme turns out to be surprisingly small~\cite{Bi}:
Only the ground state $|k\rangle$ remains, a scalar on the massless level, 
i.e. for center-of-mass momentum~$k^{\pm a}$ with $k\cdot k=0$.  The dynamics 
of this string ``excitation'' is described by a massless scalar field,
\bqr
\Phi(x)\ =\ \int\!\!d^4k\;e^{-i k\cdot x}\;\tilde{\Phi}(k) \ ,
\fqr
whose self-interactions are determined on-shell from the (amputated) string 
scattering amplitudes at tree-level,
\bqr
\langle \tilde{\Phi}(k_1)\ldots\tilde{\Phi}(k_n)
\rangle^{\rm amp}_{{\rm tree},\th} =:
A_n^{2n-4}(k_1\ldots k_n;\th) =:
\del_{k_1{+}\ldots{+}k_n}\,\tilde{A}_n^{2n-4}(k_1\ldots k_n;\th) \ .
\fqr

Interestingly, it has been shown that all tree-level $n$-point functions vanish 
on-shell, except for the three-point amplitude 
\cite{Lechtenfeld:1997za},
\bqr
\tilde{A}_3^2(k_1,k_2,k_3;\th) &=& -\frac14 \Bigl[ \,
e^{i\th} k_1^+\wedge k_2^+ -2\,k_1^+\wedge k_2^- -e^{-i\th} k_1^-\wedge k_2^-
\Bigr]^{\textstyle2}
\label{thetaamps}
\fqr
with $k_i\cdot k_j=0$ due to $\sum_n k_n=0$.  Note that $\tilde{A}_3^2$ is 
totally symmetric in all momenta.  
Expanding the square, one reads off $\tilde{A}_3^{2,M}$ for 
$M{=}{-}2,\ldots,{+}2$.  However, using the on-shell relations
\bqr
k_1^+\wedge k_2^-\ =\ h(k)^*\;k_1^+\wedge k_2^+\ =\ -h(k)\;k_1^-\wedge k_2^-
\fqr
with the phase
\bqr
h(k)\ :=\ {k^{+1} \over k^{-2}}\ =\ {k^{+2} \over k^{-1}}\ =\ 1/h(k)^*
\fqr
(identical for all three momenta), the three-point amplitude simplifies to
\bqr
\tilde{A}_3^2(k_1,k_2,k_3;\th) &=& -\frac14 [
h(k)^{1/2}\,e^{i\th/2} - h(k)^{-1/2}\,e^{-i\th/2} ]^{\textstyle4}\
( k_1^+ \wedge k_2^- )^{\textstyle2} 
\non
&=& -\frac14 \, e^{2i\th}\, [ 1 - h(k)^{-1} e^{-i\th/2} ]^{\textstyle4}\
( k_1^+ \wedge k_2^+ )^{\textstyle2} \ .
\label{threepointtheta}
\fqr
We see that the $\th$ dependence factorizes, and the contributions from
different instanton sectors differ only by powers of the leg factor~$h(k)$.
After switching to {\it real\/} $SL(2,\R)\times SL(2,\R)'$ spinor coordinates,
it is easy to see that this three-point tree-level amplitude exactly coincides, 
in the Leznov frame, with the one obtained from Plebanski's second equation
\rf{leznovframe} for the prepotential~$\psi$.  In the Yang frame, one makes
contact with Plebanski's first equation.  Furthermore, after including the 
appropriate leg factors the result becomes identical to covariant gauge 
scattering.  

The above structure of the $\th$ dependence is not a speciality of the
tree-level three-point function but actually a generic property. 
One may localize the Maxwell instantons at the worldsheet punctures and
thereby define vertex operators~$V^M$, $M{=}{-}J,\ldots,J$ 
for various instanton sectors which create an asymptotic string state
together with a Maxwell instanton out of the $M{=}0$ vacuum. 
Yet, it turns out that any two such operators are proportional to each other,
differing merely by (momentum-dependent) normalization, 
\bqr
V^M(k)\ =\ h(k)^M\;V(k) 
\label{vertexrel}
\fqr
where $V(k)$ is the vertex operator in the zero-instanton sector.
It follows that the partial amplitudes (tree or loop) in the various
instanton sectors are related by simple leg factors, and that knowledge
of a particular $A^{J,M}$ is sufficient. For this reason, we shall be
content to perform our calculations in the zero-instanton sector, except
in section four where we employ a Leznov frame.

%%%%%%%%%%%%%%%%%%%%%%%%%%%%%%%%%%%%%%%
\bigskip

\noindent
{\bf 3.2\ \ Avoiding Contact Terms}

\smallskip

The canonical computation of one-loop amplitudes entails the
use of the integrated ground state vertex operator in the $(0,0;0,0)$ 
superconformal ghost picture.  Its standard representative is
\bqr 
\tilde{V}(k) &=& \int\! d^2z\, d^2\theta d^2\bar\theta \; \exp{(ik\cdot X)} 
\\[1ex] \nonumber
&=& \int \! d^2z\,
(k^{[+}{\cdot}\partial x^{-]} -i k^-{\cdot}\psi^-\,k^-{\cdot}\psi^+)\, 
(k^{[+}{\cdot}{\bar\partial}x^{-]} +i 
k^+{\cdot}{\bar\psi}^-\,k^-{\cdot}{\bar\psi}^+) \, e^{ik\cdot x} \  . 
\label{firstvert}
\fqr 
The use of this vertex operator in amplitude calculations gives rise to 
delta functions (and squares of delta functions) on the string worldsheet
because of holomorphic/antiholomorphic Wick contractions
\bqr
\langle \partial x^{+a}(z_1)\; {\bar\partial} {\bar x}^{-b}(z_2) \rangle
\ =\ \eta^{ab} ~ \delta^{(2)}(z_1-z_2) \ .
\fqr
These contact terms are usually dropped in perturbation theory, but care
must be taken to ensure that these terms do not contribute to the scattering
in any representation.\footnote{
These contact terms are proportional, 
after the incorporation of helicity techniques, 
to inner products $\epsilon_i\cdot{\bar\epsilon}_j$ 
which vanish manifestly in the MHV amplitudes.}  
It is possible to completely avoid such contact terms by changing 
the vertex operator representative. Adding the total derivative term
\bq
-i\partial\left[ ( k^{[+}{\cdot}{\bar\partial}x^{-]} +i
k^+{\cdot}{\bar\psi}^- k^-{\cdot}{\bar\psi}^+ )  e^{ik\cdot x} \right]
-i\bar\partial\left[ ( k^{[+}{\cdot}\partial x^{-]} -i
k^-{\cdot}\psi^- k^-{\cdot}\psi^+) e^{ik\cdot x} \right]
-\partial\bar\partial e^{ik\cdot x}
\fq
and using
\bq
\partial\, e^{ik\cdot x}\ =\ k^{(+}{\cdot}\partial x^{-)}\, e^{ik\cdot x}
\fq
we arrive at
\bqr
V(k)\ =\ \int \! d^2z\,
(2k^+{\cdot}\partial x^- -i k^+{\cdot}\psi^-\,k^-{\cdot}\psi^+)\,
(2k^+{\cdot}\bar\partial x^- +i 
k^+{\cdot}{\bar\psi}^-\,k^-{\cdot}{\bar\psi}^+) \, e^{ik\cdot x} 
\label{secondvert}
\fqr
which contains $x^+$ only in the exponent and therefore precludes
not only $\langle\partial x \bar\partial x\rangle$ but also
$\langle\partial x \partial x\rangle$ and $\langle\bar\partial x 
\bar\partial x\rangle$ contractions.  In the following we shall use this 
vertex operator.  There is one drawback, however. 
Since $V(k)$ in \rf{secondvert}
is no longer invariant under complex conjugation, our computations
will not produce holomorphic squares, making chiral splitting impossible.

Next we derive the unintegrated weighted generating functional 
(Koba-Nielsen form) for $n$-point amplitudes.  The bosonic portion is  
\bq  
\prod_{j=1}^n d\theta_j d{\bar\theta}_j \int \! d\mu_n ~
 {\rm exp}\Bigl[ 
 {\smallint \! d^2z\,d^2{\tilde z}\; J^+(z) G(z,{\tilde z}) J^-({\tilde z})} 
 \Bigr]  \ .
\label{weighted}
\fq 
Here, $\theta_j$ correspond to an exponentiation
\bqr
k^+\cdot \partial x^- e^{k\cdot x}\ =\ \exp{ \left[
k\cdot x+\theta k^+\cdot \partial x^- \right] }
\Bigm|_{\rm multi-linear}
\fqr
of the pre-factor in the vertex operator from which subsequently (after
functional integration) the multi-linear part is extracted to obtain the
correlation.
For the chirally non-split form $V$ in \rf{secondvert},\footnote{
The real form $\tilde V$ of the vertex operator in \rf{secondvert} leads to
$J^-(z)$ being the complex conjugate of \rf{plusj}.}  
the currents are
\bqr  
J^+(z) &=& \sum_{j=1}^n \left[ i k_j^+ \delta^{(2)}(z{-}z_j) 
 + \theta_j k_j^+ \partial \delta^{(2)}(z{-}z_j) 
 + {\bar\theta}_j k_j^+ {\bar\partial} \delta^{(2)}(z{-}z_j) \right]
\label{plusj}
\\[1ex]
J^-(z) &=& \sum_{j=1}^n i k_j^- \delta^{(2)}(z{-}z_j) \ .
\label{minusj}
\fqr 
The sum in \rf{weighted} may be evaluated to 
\bq  
\prod_{j=1}^n d\theta_j d{\bar\theta}_j \int \! d\mu_n ~
\prod_{i\neq j} {\rm exp}\Bigl[-k_i{\cdot} k_j G_{ij} 
+ i\theta_i k_i^+ {\cdot} k_j^- \partial G_{ij} 
 + i{\bar\theta}_i k_i^+{\cdot} k_j^- {\bar\partial} G_{ij} \Bigr] 
\label{bosoncont}
\fq 
where $G_{ij} =\langle x^+(z_i,{\bar z}_i) x^-(z_j,{\bar z}_j)\rangle$, 
the bosonic two-point function on the torus, and $d\mu_n$ denotes the 
measure to integrate over the general punctured super-Riemann surface.  The 
global $N{=}2$ superspace form generalizing that in \rf{bosoncont} is 
\bq  
\prod_{j=1}^n d\theta_j d{\bar\theta}_j \int \!\! d\mu_n^s \prod_{i<j}  
{\rm exp} \Bigl[ - k_i{\cdot} k_j G_{ij} + i\theta_i k^+_i {\cdot} k^-_j D^+_i 
    G_{ij} + i{\bar\theta}_i k^+_i {\cdot} k^-_j D^-_i G_{ij}\Bigr]  ~  
 \Big\vert_{\rm multi-linear}   
\label{superform}
\fq 
where $d\mu^s_n$ is the superspace measure and $D^\pm$ the $N{=}2$
superspace derivatives.  The form in \rf{superform} is covariantized 
in the next section.  

%%%%%%%%%%%%%%%%%%%%%%%%%%%%%%%%%%%%%%%%%%%%%%%%%%
\bigskip

\noindent
{\bf 3.3\ \ Gauge Invariance and Reference Momenta}

\smallskip

In this subsection we describe the transversality of the amplitude 
at the level of the vertex operators and introduce the calculational 
tool of reference momenta \cite{ChineseMagic} in order make manifest the 
gauge invariance of the amplitudes. These instruments will allow us to compare 
the integrand with that of IIB superstring and gravity loop amplitudes.  
Spinor helicity is a useful tool in gauge theory calculations and implicitly 
has been incorporated in the $N{=}2$ string, although obscured in previous 
representations.  Here, we find it convenient to switch to a real 
$SL(2,\R)\times SL(2,\R)'$ notation 
\bqr
v^{\alpha\dot\alpha}\ =\ {1\over2}\pmatrix{
v^{+1}{+}v^{-1}{-}iv^{+2}{+}iv^{-2}&{-}iv^{+1}{+}iv^{-1}{+}v^{+2}{+}v^{-2}\cr
iv^{+1}{-}iv^{-1}{+}v^{+2}{+}v^{-2}&v^{+1}{+}v^{-1}{+}iv^{+2}{-}iv^{-2} }
\fqr
for vectors and coordinates and rewrite the $U(1,1)$ scalar product 
as~\footnote{
Note that the $\pm$ superscripts have different meaning 
on left- and right-hand sides.}
\bqr
2\,v^+\cdot w^-\ =\ \eps_{\ad\bd} \left[
v^{+\ad}w^{-\bd} - v^{-\ad}w^{+\bd} -i v^{+\ad}w^{+\bd} -i v^{-\ad}w^{-\bd}
\right] \ .
\fqr 
For a light-like momentum vector~$k^{\alp\ad}=k^\alp k^\ad$,
we have the freedom to choose the spinor $q{=}q(k)$ like~\footnote{
The matrix is degenerate and not related to the identity 
by a similarity transformation.}
\bqr
\pmatrix{ q_+ \cr q_- } = \pmatrix{ 1 & -i \cr i & 1} \pmatrix{ k_+ \cr k_- }
\qquad\qquad{\rm hence}\qquad q_+=-iq_- \ ,
\label{stringref}
\fqr
which permits us to express
\bqr
k^{+}\cdot v^{-}\ =\ -\sfrac12 q_\bet k_\bd v^{\bet\bd}
\label{canonicalnorm}
\fqr 
in $SO(2,2)$ covariant form.  Two different spinors $q_1^\alpha$ and 
$q_2^\alpha$ related to momenta $k_1$ and $k_2$ further satisfy 
\bqr 
q^+_1\ =\ {k^+_1-i k^-_1 \over k^+_2 - i k^-_2}\; q^+_2 \ .
\fqr 
A representation of the two physical polarization vectors $\eps^\pm_{\alp\bd}$
in terms of products of spinors is 
\bqr  
\epsilon^+_{\alpha\dot\beta}(k;q)\ =\ i{q_\alpha k_{\dot\beta} \over 
 q^\gamma k_\gamma} \qquad\qquad 
\epsilon^-_{\alpha\dot\beta}(k;q)\ =\ -i{q_{\dot\beta} k_\alpha \over 
 q^{\dot\gamma} k_{\dot\gamma}}
\fqr 
and has the following properties:
\bqr  
\epsilon^\pm_{\alpha\dot\beta}(k;\tilde{q})\ =\  
 \epsilon^\pm_{\alpha\dot\beta}(k;q) + f(\tilde{q},q;k)\, 
 k_{\alpha\dot\beta} \ ,
\label{referencechange}
\fqr 
\bqr  
\epsilon^+\cdot \epsilon^+ = 0 \qquad\qquad \epsilon^+\cdot \epsilon^- = -1 
 \ .
\fqr 
Because the choice of $q$ is arbitrary in any gauge-invariant 
calculation, it can be chosen to force many inner products to vanish,
considerably reducing the amount of algebra in intermediate steps of 
the calculation.

For example, in an MHV amplitude calculation the individual reference momenta
$q_i$ may be taken to coincide: $q_i=q$.  This choice eliminates all inner 
products of polarization vectors, 
\bqr  
\epsilon^+(k_1;q) \cdot \epsilon^+(k_2;q)\ =\ 0   \ . 
\label{arefidentity}
\fqr 
Since individual diagrams must, by dimensional analysis, contain
at least one inner product of two polarization vectors, the vanishing 
of the tree-level MHV (and next to MHV) amplitudes follows immediately.
Because the next-to-MHV amplitudes describe 
the self-dual scattering at tree-level, this also shows the classical 
triviality of self-dual field theory scattering \cite{Chalmers:1996rq}.  
At the loop-level it also allows a direct comparison between the N{=}2 
string amplitude calculations and those in the field theory because no 
$\partial\bar\partial G_{ij}$ arises in the integral form in \rf{superform}.  

In order to compare we normalize the $i^{\rm th}$ vertex operator 
with an additional line factor,
\bqr 
V'(k_i)\ =\ \Bigl({1\over q_i^\alpha k_{i,\alpha} }\Bigr)^2 ~ V(k_i) 
\label{normalization} 
\fqr 
with $q_i$ satisfying \rf{stringref}.  By this step, $V'$ takes the same form
as the type IIB superstring gravitational vertex operator, 
\bqr 
V'(k,\eps)\ =\ \int \! d^2z \,
\epsilon^+_{\alpha\dot\alpha}\epsilon^+_{\beta\dot\beta} 
( \partial x^{\alp\ad} -i \psi^{\alp\ad} k^-{\cdot} \psi^+ ) 
( {\bar\partial} x^{\bet\bd} +i {\bar\psi}^{\bet\bd} k^-{\cdot} {\bar\psi}^+ ) 
~e^{ik\cdot x} \ , 
\label{normalizedvertex}
\fqr 
and is clearly Lorentz covariant due to the reference momenta property in
\rf{referencechange}.  
The graviton polarization in four dimensions ($d{=}2{+}2$) is identified 
after adjoining $\epsilon^{++}_{\alpha\dot\alpha, \beta\dot\beta}(k)= 
\epsilon^+_{\alpha\dot\alpha}(k) \epsilon^+_{\beta\dot\beta}(k)$.  
Since by \rf{vertexrel} the vertex operator in a non-zero instanton
sector is related to the one in \rf{secondvert} by a leg factor only,
covariant versions of vertex operators can be given for any instanton sector
by an appropriate choice of reference momenta.

The reference momenta defined in \rf{stringref} for the different 
vertex operators satisfy 
\bqr  
q_i^\alpha q_{j,\alpha}\ =\ 0 \ , 
\fqr 
which means that this choice automatically nullifies all the different 
inner products $\epsilon^+(k_i;q_i) \cdot \epsilon^+(k_j;q_j) = 0$.  
Other choices of reference momenta, e.g. $q_j^\alpha{=}q^\alpha$ for all
external lines, may be obtained by a gauge transformation of 
the vertex operator after normalizing the external lines;
they correspond to adding a longitudinal component in \rf{referencechange}
and yield the same on-shell S-matrix elements.  

With the representation in \rf{normalizedvertex} the integrand is 
identical to the Koba-Nielsen representation of the IIB superstring, 
apart from the spin structure dependence, 
\bq 
\int\!\! d\mu_n 
\prod_{i\neq j} \exp \left(-k_i{\cdot} k_j G_{ij}\right)  
\prod_{i\neq j} \Big\vert \exp \Bigl[  
\epsilon_{[i}{\cdot}k_{j]} \partial_i G_{ij} + 
\epsilon_i{\cdot}\epsilon_j \partial_i\partial_j G_{ij} + 
\epsilon_i{\cdot}\bar\epsilon_j \partial_i \bar\partial_j G_{ij} 
\Bigr] \Big\vert^2_{\rm multi-linear} 
\label{ampepsilon}
\fq 
where the label `multi-linear' means that the integrand is expanded 
in powers of the polarizations, keeping only the terms 
linear in each polarization ($\eps_j$ or $\bar\eps_j$).
The $N{=}1$ superspace form has 
\bqr
\partial_i G_{ij} \rightarrow D_+^i G_{ij} \qquad
\partial_i\partial_j G_{ij} \rightarrow D_+^i D_+^j G_{ij} \qquad
\partial_i \bar\partial_j G_{ij} \rightarrow D_+^i D_-^j G_{ij} \ .  
\fqr
This procedure accounts for the $\theta$ integrations in the preceeding 
form in \rf{bosoncont}, after choosing the reference momenta such that
all $\epsilon_i\cdot\epsilon_j=0$, $\epsilon_i\cdot\bar\epsilon_j =0$ and
$\bar\epsilon_i\cdot\bar\epsilon_j=0$.  The reference momenta that occur 
naturally in the vertex operator for the $N{=}2$ string in \rf{stringref} 
force all inner products in \rf{ampepsilon} $\epsilon_i\cdot\epsilon_j=0$ 
and $\epsilon_i \cdot{\bar\epsilon}_j=0$ via \rf{arefidentity} and we 
regain \rf{superform}, although an arbitrary choice of $q_i$ demonstrates 
the covariance in \rf{ampepsilon}.

\section{Three-point Genus One} 

In this section we calculate the genus-one closed-string three-point 
amplitude originally derived (the $M{=}0$ part) in \cite{Bonini:1991bf} 
and compare the result with field theory, i.e. self-dual gravity. 
As mentioned in the previous section, in the Leznov frame
the tree-level expression $A_3^{J=2}(\th{=}0)$ from \rf{threepointtheta}
exactly produces the field-theory result generated from the Lagrangians
\rf{secondplebanski} or \rf{twofield2},
\bqr
A_3^{J=2}(\th{=}0)\ =\ A_3^{\rm tree}\ =\ 
( \eps_{\ad\bd}\, k_1^{+\ad}\,k_2^{+\bd} )^2 
\fqr
where we switched to real spinor notation again.  
Other formulations of self-dual gravity are related by appropriately 
normalizing the external lines.  In the gauge 
choice of \rf{leznovframe} and without the external line factors required for 
covariance, the field-theoretic one-loop expression $A_3^{\rm 1-loop}$ is, by 
dimensional analysis, constrained to be
\bqr
A_3^{\rm 1-loop}\ =\ (k_1^{+\ad}\,k_{2\,\ad}^+)^6 \, \bar{A}_3^{\rm SDG} \ .
\label{3tensor}
\fqr
This fixes the tensor structure.
The remaining proportionality factor~$\bar{A}_3^{\rm SDG}$ in the amplitude 
then boils down to a field-theoretic triangle integral.

The triangle integrals appearing below are infra-red divergent as on-shell 
kinematics require $k_i^2=k_i^+\cdot k_i^-=0$.  
The field-theory loop calculation 
can also be performed by keeping $k_3^2\neq 0$ until after the integration, 
which generates the infra-red divergence as $k_3^2\rightarrow 0$.   
Direct comparison with the on-shell string scattering is independent 
of this limit. 

After introducing Feynman parameters and Schwinger time,  
the three-point on-shell one-loop amplitude becomes
\bqr  
\bar{A}_3^{\rm 1-loop} &=& \int {d^d\ell\over (2\pi)^d} 
\int_0^\infty dT ~ T^2 ~ 
\int_0^1 da_1 da_2 da_3 ~ \delta( 1- a_1-a_2 -a_3) 
\non && \times\, 
\exp\Bigl[-T\left(a_1 \ell^2+a_2(\ell-k_1)^2+a_3(\ell+k_3)^2\right)\Bigr] ~ 
\non && \times\, 
(\ell^{+\ad}\,k_{1\,\ad}^+)^2 \left( (\ell-k_1)^{+\ad}\,k_{2\,\ad}^+\right)^2
\left( \ell^{+\ad}\,k_{3\,\ad}^+\right)^2  
\fqr 
which, after shifting 
\bqr  
\ell\ =\ \ell' + a_2 k_1 - a_3 k_3 \ , 
\fqr 
takes the form of \rf{3tensor}, with
\bqr  
\bar{A}_3^{\rm SDG}\ =\ \int {d^d\ell\over (2\pi)^d} \int_0^\infty dT ~ T^2 
\int_0^1 da_1 da_2 da_3 ~ a_1^2 a_2^2 a_3^2~ \delta(1-\sum_{j=1}^3 a_j)  
\exp\left[-T\ell^2\right]  \ . 
\fqr 
Integrating over the loop momentum in (unregulated) $d{=}4$ real 
dimensions and restoring the tensor structure gives
\bqr 
A_3^{\rm 1-loop}\ =\ (k_1^{+\ad}\,k_{2\,\ad}^+)^6 
\times {1\over 16 \pi^2} \times {1\over 15 \times 5!} \int dT \ . 
\label{3ptfield}
\fqr 
The integral is IR divergent,\footnote{
It vanishes in dimensional reduction or regularization.}
and we regulate it by imposing
a  Schwinger proper-time cutoff at $T=T_{\rm max}$; 
the unregulated results for both the field theory and string theory 
may be compared without referring to a regulator.     

The three-point function in \rf{3ptfield} is to be compared with
the $N{=}2$ string result found next.  The string-theory calculation in 
the Leznov frame confirms the same tensor structure as in the field theory, 
\bqr
A_3^{J=6}(\th{=}0) \ =\ (k_1^{+\ad}\,k_{2\,\ad}^+)^6 \, \bar{A}_3^{N=2}
\fqr
which differs from the three-point scattering found in \cite{Bonini:1991bf}
only by normalization ($\th{=}0$ instead of $M{=}0$).  We may therefore
take over their result, 
\bqr
\bar{A}_3^{N=2}\ =\ \int\!{d^2\tau\over \tau_2^2}\, E_3(\tau,\bar\tau) \ ,
\label{stringform}
\fqr 
where the non-holomorphic Eisenstein series is defined as
\bqr  
E_3(\tau,\bar\tau)\ = 
\sum_{(m,n)\neq (0,0)} {\tau_2^3\over \vert m+n\tau\vert^6} 
\fqr 
and satisfies 
\bqr  
\tau_2^2 \partial_\tau \partial_{\bar\tau} E_3(\tau,\bar\tau) 
\ =\ 6 E_3(\tau,\bar\tau) \ .
\label{diff}
\fqr 
Using \rf{diff} the integral in \rf{stringform} can be evaluated, 
\bqr  
\bar{A}_3^{N=2}\ =\
\int\!{d^2\tau}\,(\partial^2_{\tau_1} + \partial^2_{\tau_2}) E_3(\tau,\bar\tau) 
\ =\ \sfrac16 \partial_{\tau_2} E_3(\tau,\bar\tau) \vert_{\tau_2=\kappa} 
\label{stringform2}
\fqr 
where the integral has been regulated by cutting its large-$\tau_2$ region
at $\tau_2=\kappa$.  
The over the boundary term at the small-$\tau_2$ end of the 
fundamental keyhole domain 
\bqr  
\vert\tau\vert\geq 1 \qquad{\rm and}\qquad
\vert\tau_1\vert\leq {1\over 2} 
\fqr  
is zero.  For $\tau_2\to\infty$, $E_3$ has the asymptotic form 
\bqr  
E_3\ =\ 2\,\zeta(6)\,\tau_2^3 + \sqrt{\pi} \zeta(5)\,\Gamma(5/2)\,\tau_2^{-2} 
+ O(e^{-2\pi\tau_2})  \ ,  
\fqr 
yielding for \rf{stringform2} the expression 
\bqr  
\bar{A}_3^{N=2}\ =\,\zeta(6)\,\tau_2^2 \vert_{\kappa} 
\fqr 
together with terms that vanish as $\kappa\rightarrow\infty$.  
Bringing back the tensor structure, we end up with
\bqr
A_3^{J=6}(\th{=}0)\ =\ (k_1^{+\ad}\,k_{2\,\ad}^+)^6 \;
\,\zeta(6)\,\tau_2^2 \vert_{\kappa} \ ,
\label{3ptN2amp}
\fqr
with $\zeta(6)= \pi^6/945$.  

The three-point functions in \rf{3ptfield} and in \rf{3ptN2amp} agree 
after redefining the string proper time $\tau_2^2=T$.  The regulator
\bqr
T_{\rm max}\ =\ \kappa^2 
\label{regulator}
\fqr 
together with a normalization that can be absorbed in \rf{regulator}  
gives the match.   

The two integrals \rf{3ptfield} and \rf{stringform} differ 
by a factor of $\tau_2$ or, in the field theory interpretation, 
a shift in dimension~\cite{Marcus}.  
The matching of the scattering at three-point order is simply a redefinition 
of the Schwinger proper-time or the cutoff.  This is inconsequential at 
three-point order because the results are both infra-red divergent. 
However, for the finite higher-point amplitudes such a redefinition 
is not possible, and the mismatch by a factor of $\tau_2$ makes for a crucial 
difference between the two theories.

%%%%%%%%%%%%%%%%%%%%%%%%%%%%%%%%%%%%%%%%%%%%%%%%%%%%%%%%%%%%%%%%%%%%%%%%%%

\section{Four-Point Genus One} 

\noindent
{\bf 5.1\ \ String Integrand}

\smallskip

In this section we analyze the measure for the integration of the
four-point (and higher-point) amplitudes for the $N{=}2$ closed string in
the critical dimension $d{=}2{+}2$ and compute the integrand in terms of the
bosonic and fermionic worldsheet correlators.  The field-theory limit
is taken in order to compare with the self-dual field theory and one-loop
maximally helicity violating amplitudes in gravity.  The comparison between
the measure factors in the string and field theory persists to multi-genus.

The $N{=}2$ superconformal algebra has as its generators the energy-momentum
tensor $T$, two supercurrents $G^\pm$, and the $U(1)$ current $J$.  
The associated ghost structure consists of the $(b,c)$ diffeomorphism ghosts, 
the $(\bet^\mp,\gam^\pm)$ local supersymmetry ghosts, 
and an additional $(b',c')$ ghost system 
for the local $U(1)$ invariance or R symmetry.
Each chiral $N{=}2$ matter multiplet $X=(x,\psi)$ and each ghost system
contributes a (modular invariant) determinant factor to the one-loop string 
integration measure (continued to $d$ dimensional target spacetime)
\bqr
Z_d\ab(\tau,\bar\tau)\ =\ 
Z_x(\tau,\bar\tau)\, Z_{\psi}\ab(\tau,\bar\tau)\, Z_{bc}(\tau,\bar\tau)\,
Z_{\beta\gamma}\ab(\tau,\bar\tau)\, Z_{b'c'}(\tau,\bar\tau)
\ ,
\fqr
with the respective factors being 
\bqr
Z_x(\tau,\bar\tau)\ =\ \tau_2^{-d/2} \vert\eta(\tau)\vert^{-2d} \qquad\qquad
Z_\psi\ab(\tau,\bar\tau)\ =\ 
\vert \vartheta\ab(0,\tau)\vert^d\, \vert \eta(\tau) \vert^{-d} \ ,
\label{matterfields}
\fqr
\bqr
Z_{bc} (\tau,\bar\tau)\ =\ \tau_2\,\vert \eta(\tau)\vert^4 \qquad\qquad
Z_{\beta\gamma}\ab(\tau,\bar\tau)\ =\
\vert \vartheta\ab(0,\tau)\vert^{-4}\, \vert \eta(\tau) \vert^4
\fqr
and, for the one associated with the local $U(1)$ symmetry, 
\bqr
Z_{b'c'}(\tau,\bar\tau)\ =\ \tau_2\,\vert \eta(\tau)\vert^4  \ .
\fqr
The building blocks are the Jacobi theta functions (featured in the Appendix)
with continuous characteristic $\ab$ equal to spin structure 
and the Dedekind eta function 
\bqr
\eta(\tau)\ =\ q^{1/24} \prod_{n\neq 0} (1-q^n)  
\qquad{\rm where}\quad q=e^{2\pi i\tau} \ ,
\fqr
with $\tau$ denoting the modular parameter of the torus.
For general $d$ the product of all determinant factors combines into
\bqr
Z_d\ab(\tau,\bar\tau)\ =\ \tau_2^{-{(d-4)\over 2}}\, 
\vert \vartheta\ab(0,\tau)\vert^{d-4}\,
\vert \eta(\tau)\vert^{-3(d-4)} \ ,
\label{partitionfunction}
\fqr
and equals unity in four real dimensions~\cite{MM}. This point trivializes 
the spin structure summation for the one-loop partition function and signals 
the absence of a tachyonic mode otherwise arising from the $q$-expansion 
of eta functions.

Superconformal gauge fixing of the worldsheet $N{=}(2,2)$ supergravity
produces not only constraints and their ghost systems but also reduces
the supergravity path integral to one over the associated finite-dimensional
moduli spaces. After explicitly performing the fermionic moduli integrals,
which generate picture-raising insertions, one is left with reparametrization
and Maxwell moduli. Both come in two varieties: moduli encoding the shape of
the $U(1)$ bundle over the worldsheet, and moduli describing the locations
and $U(1)$ monodromies of the vertex operators. In the torus case, the former
are $(\tau,\bar\tau)$ and $\ab$ while the latter comprise $\{(z_i,\bar z_i)\}$ 
and twist angles $\{(\rho_i,\bar\rho_i)\}$ interpolating between 
NS- and R-type puncture.\footnote{
Isometries fix the reparametrization and 
Maxwell moduli of one of the punctures.} 
Since for genus one the Jacobian torus of spin structures is isomorphic to the
worldsheet itself we may parametrize it by an additional torus variable,
\bqr
u\ =\ (\sfrac12{-}\alpha )\, \tau + (\sfrac12{-}\beta) \ .
\fqr
The modular invariant integration measures are
\bqr
{d^2\tau\over\tau_2^2} \qquad\qquad{\rm and}\qquad\qquad
{d^2u\over\tau_2}
\label{spinmeasure}
\fqr
on the fundamental domain~$\cF$ of $PSL(2,\Z)$ and the torus~$\cT$,
respectively. Due to spectral flow, the integrand is independent of the
twist angles, whose integration thus results merely in a 
constant volume factor for each puncture. The integration over the
puncture locations, however, are nontrivial but modular invariant in
the combination $\int_\cT d^2z\,V(z,\bar z)$.

Putting everything together, the scattering amplitude is given by
\bqr
A_n(k_j)\ =\ 
\int_{\cF} \!{d^2\tau\over\tau_2^2} \int_{\cT} \!{d^2u\over\tau_2}~
\prod_{j=1}^{n} \int_{\cT} \!d^2z_j  ~
\prod_{i < j} e^{-k_i\cdot k_j G_{ij}} ~
K_{KN}(z_i,{\bar z}_i;u,\bar{u};\tau,\bar\tau) \ ,
\label{KNform}
\fqr
with $K_{KN}$ labeling the contractions between the vertex fields.

The expansion of $K_{KN}$ has a suggestive form after the grouping of terms
that we now turn to. Each term with fermionic contractions can be paired
with a purely bosonic term.  This property is a consequence of worldsheet 
$N{=}2$ superconformal invariance and can also be used to prove the
vanishing of the corresponding tree-level amplitudes.  

We break the contractions into three groups of terms and analyze 
the contributions from the string scattering when the reference momenta 
are chosen to agree with \rf{stringref}.  
These holomorphic and anti-holomorphic 
mirror terms are depicted graphically in Figures 1 and 2.  
The bosonic propagator is
\bqr
G_{ij}\ =\ -\ln E(z_i{-}z_j) - \ln E({\bar z}_i{-}{\bar z}_j)
 + {2\pi\over\tau_2} \Bigl[{\rm Im}(z_i{-}z_j)\Bigr]^2
\label{bosonprop}
\fqr
where $E$ is the prime form on the torus,
\bqr
E(z,\tau)\ =\ {\vartheta\hh (z,\tau)
  \over \vartheta'\hh (0,\tau)} \ ,
\fqr
and the latter term in \rf{bosonprop} subtracts the bosonic zero mode from
the kernel.\footnote{
Our convention is that $G_{ij}$ marks the full propagator 
including holomorphic, anti-holomorphic and zero-mode term; 
the holomorphic piece, $-\ln E(z_i{-}z_j)$, will be denoted by $G(z_{ij})$, 
explicitly displaying the holomorphic coordinate.}
The holomorphic half of the fermionic propagator is the Szeg\"o kernel 
for general continuous monodromies,
\bqr
S\ab(z,\tau)\ =\
{ \vartheta\ab(z,\tau)\, \vartheta'\hh(0,\tau) \over
  \vartheta\ab(0,\tau)\, \vartheta\hh(z,\tau) } \ ,
\fqr
except for the $\alpha{=}\beta{=}1/2$ periodic sector in which an additional
zero mode develops.  Expansions of the propagators are given in the Appendix.

The first type of term in $K_{KN}$ is
\bqr
I^{(1234)} &=&  
k_1^+\cdot k_2^-\, k_2^+ \cdot k_3^-\, k_3^+\cdot k_4^-\, k_4^+ \cdot k_1^-
\non
&&\times\,
\Bigl( \partial_1 G_{12} \partial_2 G_{23}
 \partial_3 G_{34}\partial_4 G_{41} -
  S_{12}\ab S_{23}\ab S_{34}\ab S_{41}\ab \Bigr) \ .
\label{termone}
\fqr
Its reverse ordering $(4321)$ is denoted by $I^{(4321)}$.  The
latter gives the complex conjugated contribution via $k^+\leftrightarrow
k^-$.  In addition we need the remaining permutations, $I^{(1324)}$,
$I^{(1243)}$, and their reverse orderings.  This set is closed under
permutation of any two indices.

\begin{figure}[htb]
\vskip.5cm
\begin{center}
\begin{minipage}{6cm}
\begin{center}
\epsfig{file=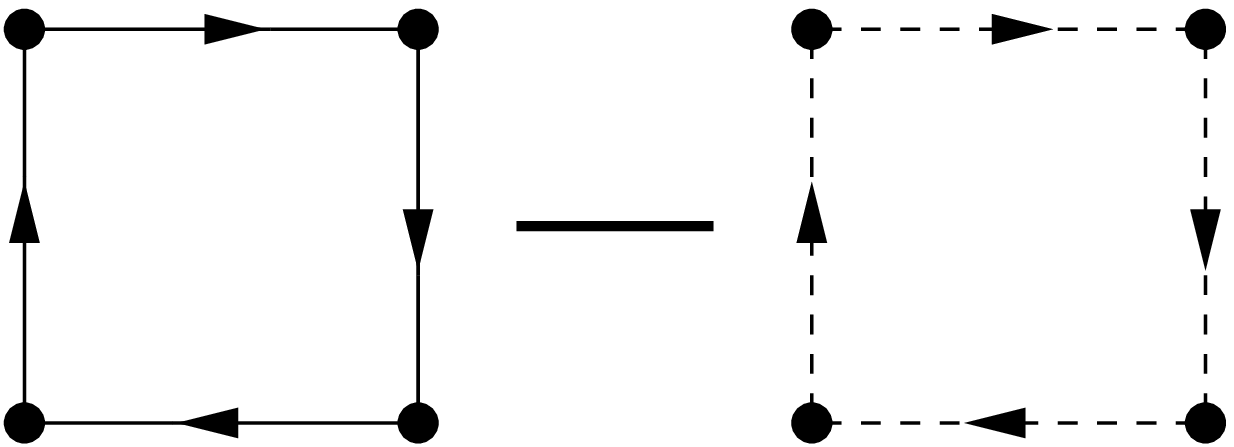,height=2cm}\\
(a)
\end{center}
\end{minipage} \hskip .3in 
\begin{minipage}{6cm}
\begin{center}
\epsfig{file=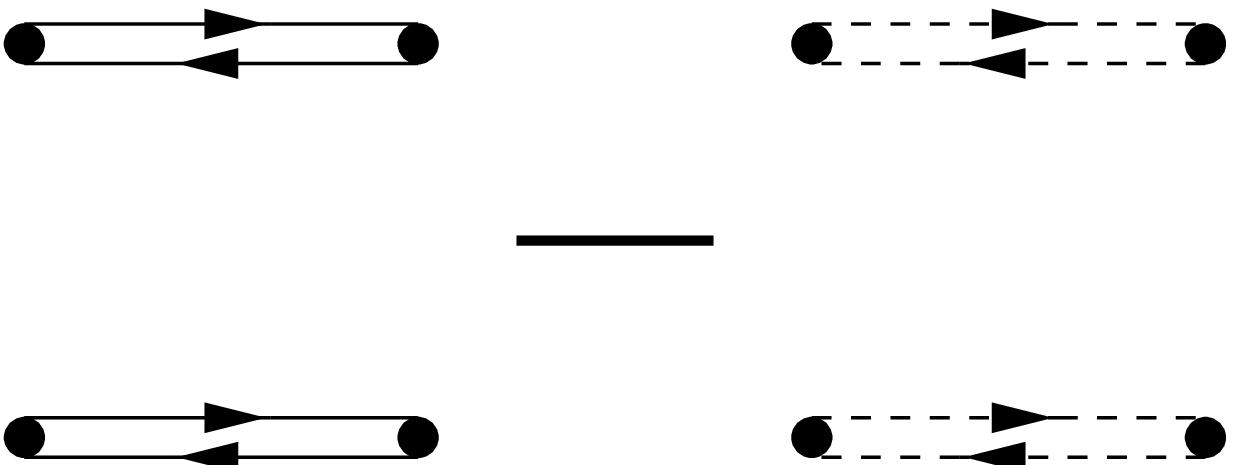,height=2cm}\\
(b)
\end{center}
\end{minipage}
\end{center}
\vskip-.5cm
\caption{\sl
Contributions (a) $I^{(1234)}$ and (b) $I^{(12)(34)}$
to the $N{=}2$ closed string amplitude.  The solid
lines are derivatives of the bosonic two-point correlator and the dashed
lines are holomorphic (or anti-holomorphic) fermionic Green's functions.}
\vskip.2cm
\end{figure}

Next we have the three terms
\bqr
I^{(12)(34)} &=& 
- k_1^+\cdot k_2^-\, k_2^+\cdot k_1^-\, k_3^+\cdot k_4^-\, k_4^+\cdot k_3^-
\non
&&\times\,
\Bigl( \partial_1 G_{12}
 \partial_2 G_{21} \partial_3 G_{34} \partial_4 G_{43} -
 S_{12}\ab S_{21}\ab S_{34}\ab S_{43}\ab \Bigr) \ ,
\label{termtwo}
\fqr
together with the orderings $I^{(14)(23)}$ and $I^{(24)(13)}$.  The
terms in \rf{termtwo} are products of pairs of Szeg\"o kernels as
opposed to the cyclic combinations in \rf{termone}.

The remaining terms are paired so that there are products of only
two Szeg\"o kernels (in a cyclic fashion),
\bqr
I^{(12)} &=& 
k_1^+\cdot k_2^-\, k_2^+\cdot k_1^- 
\Bigl( \partial_1 G_{12} \partial_2 G_{21} - S_{12}\ab S_{21}\ab \Bigr)
\non
&&\times\,
\Bigl( k_3^+\cdot k_1^-\, \partial_3 G_{31} + 
       k_3^+\cdot k_2^-\, \partial_3 G_{32} + 
       k_3^+\cdot k_4^-\, \partial_3 G_{34} \Bigr)
\non
&&\times\,
\Bigl( k_4^+\cdot k_1^-\, \partial_4 G_{41} + 
       k_4^+\cdot k_2^-\, \partial_4 G_{42} + 
       k_4^+\cdot k_3^-\, \partial_4 G_{43} \Bigr) \ ,
\label{termthree}
\fqr
together with its permutations: $I^{(34)}$, $I^{(14)}$,
$I^{(23)}$, $I^{(24)}$, and $I^{(13)}$.   Terms with
three fermion pairs contracted cancel.

\begin{figure}[htb]
\vskip.5cm
\begin{center}
\epsfig{file=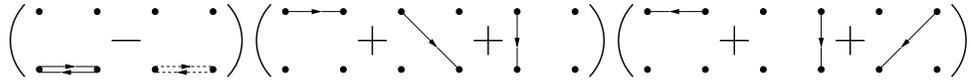,height=1cm}
\end{center}
\vskip-.5cm
\caption{\sl
Additional contributions $I^{(34)}$ to the $N{=}2$ closed string amplitude.}
\vskip.2cm
\end{figure}

The gauge-invariant vertex operators normalized as in \rf{ampepsilon}
produce the same set of terms as in \rf{termone}, \rf{termtwo} and
\rf{termthree} but with the modification
\bqr
k_i^+\cdot k_j^- \ \to\ \eps_i\cdot k_j
\fqr
everywhere.  Before and after integrating over spin structures, and with the 
choice of reference momenta $q_i=q$, this substitution shows that the 
zero-slope limit of the closed-string amplitude reproduces the Feynman
rules of gravity one-loop amplitudes without any $\epsilon_i\cdot
\epsilon_j$ or $\epsilon_i\cdot\bar\epsilon_j$ terms (i.e. MHV structure).

Let us analyze the structure of $K_{KN}$ in \rf{KNform} 
given the boson/fermion pairing of the various terms in the expansion.  
For the periodic spin structure $\hh$, 
\bqr
S_{ij}\ab\ \to\ \partial_i\,G_{ij}
\fqr
for each Szeg\"o kernel, and each set of terms in eqs. \rf{termone}, 
\rf{termtwo} and \rf{termthree} vanishes identically.  Furthermore, at generic 
values of $\ab$ the integrand vanishes at coincident points $z_i{-}z_j\to0$, 
making contact with the vanishing tree-level result 
via worldsheet degeneration. 
More explicitly, in the short-distance limit of coincident points one gets
\bqr
G(z_i{-}z_j)\ =\ - \ln(z_i{-}z_j) \quad,\qquad
S(z_i{-}z_j)\ =\ {1\over z_i{-}z_j}\ =\ -\partial\,G(z_i{-}z_j) \ , 
\fqr
and the integrand is zero pointwise before integration
over the vertex operators.  This cancellation can be explained 
in a number of ways.  First, in field theory this is due to the
fact that every tree diagram in gauge theory (Yang-Mills or gravity) contains 
at least one contraction $\epsilon_i\cdot\epsilon_j$, and the identical 
reference momenta choice for all external lines in an MHV helicity 
configuration nullifies these terms. 
Second, at tree-level, target spacetime supersymmetric 
Ward identities in a supersymmetric gauge theory force the MHV 
amplitudes to be identically zero (in a supersymmetric gauge theory 
the tree-level graviton or gauge theory scattering amplitude does 
not contain internal fermion lines). Third, although the $N{=}2$ string
is not spacetime supersymmetric, the worldsheet $N{=}2$ superconformal 
invariance of the vertex operator
forces the tree-level amplitude to be zero. 

%%%%%%%%%%%%%%%%%%%%%%%%%%%%%%%%%%%%%%%%%%%%%%%%%%%%%%
\bigskip

\noindent
{\bf 5.2\ \ Comparison with Field Theory at Zero-slope}

\smallskip

In this subsection we take the zero-slope limit of the amplitude obtained
in the previous subsection and compare it with the field-theory computation
obtained in self-dual gravity at one-loop \rf{gravfourrelation}.   
Since the integration over the spin structures may be performed before
or after the $\alp\to0$ limit and it is not a priori obvious whether the
ordering matters (because of singularities at the periodic spin structure),
we will examine both orderings: field-theory limit first in the present
section, spin structure integration first in the next one.
The results will turn out to be the same.

The amplitude from the string differs from self-dual gravity amplitudes
in $d{=}2{+}2$ because of the $(b',c')$ ghost system associated to
the $U(1)$ R symmetry. Thanks to it, the integrand contains
an additional $\tau_2$ factor when compared to the integrand
of type IIB superstring theory projected
onto the self-dual sector of gravity in four dimensions (for example,
by toroidal compactification on $T^6$ to the non-supersymmetric sector).
Quite generally, a  factor of~\footnote{
The $n$ factors of $\tau_2$, one for every vertex operator, arise from
the mapping of the torus to the unit square by $z_i=x_i+\tau y_i$, with
$x_i,y_i \in [0,1]$.}
\bqr
\int\!{d\tau_2\over \tau_2} ~ \tau_2^{n-d/2} ~ e^{-\tau_2 f(k_i)}
\label{taufactor}
\fqr
is associated with writing an $n$-point $\phi^3$ Feynman diagram 
in $d$ dimensions as
\bq
\int\!\!{d^d\ell\over (2\pi)^d} \prod_{j=1}^n {1\over (\ell{-}p_j)^2} =
  (4\pi)^{-d/2} \Bigl(\prod_{j=1}^n \int_0^1\!\! da_j \Bigr)
  \delta(1-\Sigma_{j=1}^n a_j) 
  \int_0^\infty\!\! dT ~ T^{n-1-d/2}\, e^{-T f(k_i,a_i)} 
\fq
where $f(k_i,a_i)=-(\sum_k p_ka_k)^2+\sum_k p_k^2a_k$.

The field-theory limit of the string amplitude is obtained by transforming
the string worldsheet coordinates for the vertex operators into a Schwinger
proper-time form.  From the field-theory point of view, the higher-$q$ terms 
correspond to the exchange of massive modes (which are absent in the $N{=}2$ 
string).  We briefly examine the full analytic structure in the limit.  
Following \cite{D'Hoker:1995yr,Chalmers:1998dc} in the analytic extraction of 
poles, we introduce new variables $w_{ij}$ satisfying 
$\vert w_{ij} \vert \leq 1$ and defined by
\bqr
 w_{ij}\ =\ \left\{ \begin{array}{lr}
e^{2\pi i z_{ij}} & \qquad {\rm for} \quad {\rm Im}~ z_{ij} > 0 \\
q\,e^{2\pi i z_{ij}} & \qquad {\rm for} \quad {\rm Im}~ z_{ij} < 0
\end{array} \right. 
\label{wvariables}
\fqr
with $z_{ij}\equiv z_i{-}z_j$.
We also make use of the standard parametrization of the vertex insertion
points in terms of the real variables $\alpha _i$ and $u_i$
\bqr
\begin{array}{ll}
u_1\ =\ y_1 \qquad\quad & \alpha_1\ =\ 2\pi(x_1+u_1\tau_1) \\
u_2\ =\ y_2-y_1 \qquad\quad & \alpha_2\ =\ 2\pi(x_2-x_1+u_2\tau_1) \\
u_3\ =\ y_3-y_2 \qquad\quad & \alpha_3\ =\ 2\pi(x_3-x_2+u_3\tau_1) \\
u_4\ =\ 1-y_3 \qquad\quad & \alpha_4\ =\ 2\pi\tau_1-\alpha_1-\alpha_2-\alpha_3
\end{array} \ ,
\label{reparametrization}
\fqr
where $u_1+u_2+u_3+u_4=1$ and $\alp_1+\alp_2+\alp_3+\alp_4=2\pi\tau_1$.
This can be achieved by using the translational symmetry of the torus to 
fix the position of one vertex operator insertion point.

Since only logarithmic derivatives of the prime form multiply the
Koba-Nielsen term, multiplying $E$ with a $z$-independent factor produces 
only a constant shift in the Koba-Nielsen exponent which vanishes as a result 
of momentum conservation.  We are therefore entitled to neglect constant 
factors in $E$ and simplify its product representation,
\bqr
E(z_{ij})\ =\ {\vartheta\hh(z_{ij},\tau)
  \over \vartheta'\hh(0,\tau)}
\ \ {\dot =}\ \ e^{\pi i z_{ij}} 
 \prod_{n=0}^{\infty} \bigl( 1- q^n e^{-2\pi i z_{ij}} \bigr)
 \bigl( 1- q^{n+1} e^{2\pi i z_{ij}} \bigr)  
\label{primeformexpand}
\fqr 
and we define
\bqr
{\cal R}(w_{ij})
\ =\ \prod_{i\neq j}\prod_{n=0}^{\infty}\vert 1-w_{ij}q^n\vert^{-s_{ij}} \,
\label{Rfunctor}
\fqr
which is the component of $e^{\half s_{ij} G(z_{ij})}$ that contains all 
the infinite products from the expansion in \rf{primeformexpand}.
The remaining contributions from the Koba-Nielsen terms
$\prod_{i<j}e^{\half s_{ij} G(z_{ij})}$ that stem from the $e^{\pi i z_{ij}}$
part of the prime forms and from the zero-mode subtractions in the bosonic
Green's functions can be combined into the expression
$|q|^{-(su_1u_3+tu_2u_4)}$.

The full amplitude for a given spin structure $\ab$ may then be rewritten as
$A_4\ab(s,t)+A_4\ab(t,u)+A_4\ab(u,s)$, with
\goodbreak
\bqr
A_4\ab(s,t) &=&
\int_{\cal F} \! d^2\tau ~\tau_2^2~
\prod^{4}_{i=1} \int_0^{2\pi} \! {d\alpha_i \over 2\pi}  ~
  \delta (2\pi\tau_1 - \Sigma_j \alpha_j)
\cr &&
\times \prod_{i=1}^4 \int_0^1 \! du_i ~\delta (1-\Sigma_j u_j) ~
\vert q \vert^{-(s u_1 u_3 + t u_2 u_4)}~{\cal R}(w_{ij}) ~ K_{KN} 
\label{amplitudeexpand}
\fqr
and the obvious permutations.  We note that, for a given spin structure, 
$K_{KN}$ is identical to the MHV kinematic factor in IIB superstring theory 
in \rf{ampepsilon}.  The function ${\cal R}$, defined from the product 
expansion of the $\vartheta$-functions as in \rf{Rfunctor}, may be expanded 
in an infinite series as follows:
\bqr
{\cal R}(w_{ij})\ =\
 \prod _{i=1} ^4 \bigl | 1 - e^{i \alpha _i} |q|^{u_i}
\bigr | ^{-s_i} \sum _{n_i=0} ^\infty \sum _{|\nu _i|\leq n_i} P^{(4)}
_{\{n_i \nu _i\}} (s,t)
\prod _{i=1} ^4 |q|^{n_i u_i} e^{i \nu _i \alpha _i}  \ .
\label{breakup}
\fqr
Here, $s_i=s$ for $i$ even, $s_i=t$ for $i$ odd, and $P^{(4)}_{\{n_i\nu_i\}}
(s,t) $ are polynomials in $s$ and $t$ that may be generated recursively.
Consider now the identity
\bqr
\int_0^{2\pi} \!\frac{d\alpha}{2\pi} ~ e^{i\eta\alpha}
\left| 1-x e^{i\alpha} \right|^{-s}
\ =\ x^{-r} \int_0^\infty \! d\beta ~ x^\beta\,\varphi_{r\eta}(s;\beta) \ ,
\fqr
where
\bqr
\varphi_{r\eta}(s;\beta)\ =\ \sum_{k=0}^\infty C_k(s)\,C_{k+|\eta|}(s)\,
\delta(2k{+}r{+}|\eta|{-}\beta)
\fqr
is the inverse Laplace transform of a hypergeometric function, and
\bqr
C_k(s)\ =\ \frac{\Gamma(\frac{s}{2}+k)}{\Gamma(\frac{s}{2})\,\Gamma(k+1)} \ .
\fqr
For $x$ being some power of $|q|$ we can take advantage of this identity
and execute the integration over the angular variables $\alpha_i$.
We observe that the zero-slope limit is equivalent to putting 
${\cal R}(w_{ij})=1$ from the start, since the effect of a nontrivial
function ${\cal R}$ is felt only at higher order in $\alp'$.
The analysis can be extended to contain the angular parameters associated
with the kinematical factor $K_{KN}$ and justifies the substitution rules
that follow from the closed-string context.  The remaining propagator terms
in $K_{KN}$ generate the Feynman parameters associated with the derivative
couplings of the field-theory vertex in the zero-slope limit; the kinematical
expression is identical to that obtained in the IIB superstring before
summing over spin structures.  This procedure has been systematized at
one-loop and $n$-point through the Bern-Kosower string-motivated rules for
calculating gauge theory scattering \cite{Bern:1988tw} adapted to gravity
\cite{Bern:1993wt}.

If it were not for the non-holomorphic zero-mode part in \rf{bosonprop},
perfect bose-fermi cancellation in eqs. \rf{termone}, \rf{termtwo}, and
\rf{termthree} would occur in the field-theory limit.
As it is, however, the remainder of the various pairings of the 
bosonic contractions with the fermionic ones
is proportional to at least one factor of $(2\pi/i\tau_2)
{\rm Im}\,z_{ij}$.  The zero modes explicitly break the holomorphicity
of the string scattering, and the MHV amplitude may be
understood as a holomorphic anomaly in the zero-slope limit of the
$N{=}2$ string. 

Remaining in \rf{amplitudeexpand} is the single factor of $\tau_2^2$
multiplied by bosonic zero-mode contributions from $K_{KN}$.
Comparing with \rf{taufactor} we see that this corresponds to a 
field-theory result in $d{=}2$ real dimensions~\cite{Marcus}.
The $\tau_2$-dependence in the low-energy scattering
of gravitons and the $d{=}2$ technical interpretation follows from
either simple toroidal compactifications of the IIB superstring, 
or using the well-established string-inspired Feynman rules adapted
to the case of perturbative gravity \cite{Bern:1993wt}. The latter we
briefly discuss next in order to map the kinematical structure $K_{KN}$
to the MHV one-loop gravity amplitudes.

The string-inspired generation of the graviton scattering amplitude, 
which in the zero-slope limit arises from the corners of the moduli 
space, involves the Feynman-parametrized form originating from a 
$\phi^3$ diagram \cite{Bern:1993wt}
\bqr
D\ =\ c_n\int_0^1\!dx_{i_{n-1}}\ldots\int_0^{x_{i_2}}\!dx_{i_1}{K_{\rm red}
\over \left( \sum_{a<b}^n P_{i_a}{\cdot} P_{j_b}\, x_{{i_a}{j_b}}\,
(1{-}x_{{i_a}{j_b}}) \right)^{n-d/2}} \ ,
\label{multilinear}
\fqr 
where, in $d$ dimensions, 
\bqr  
c_n\ =\ {\left(4\pi\right)^{2-d/2}}\,{\Gamma(n{-}d/2)\over 16\pi^2} \ . 
\fqr 
$P_{i}$ is the momentum flowing into the $i^{\rm th}$ leg of the
$n$-gon $\phi^3$ diagram, $x_{ij}{=}x_i{-}x_j$,
and $x_i$ are Feynman parameters.  All $\phi^3$
diagrams are to be considered  with external trees attached where
the external lines follow a cycle
ordering; in the gravitational case we sum over all of the non-cyclic
orderings without associated Yang-Mills color factors.  
The factor $K_{\rm red}$ comprises terms generated from the kinematical 
expression identical to the Koba-Nielsen term of the multi-graviton 
scattering amplitude,
\bqr
\left. \exp{\left[(k_i\cdot\epsilon_j-\epsilon_j\cdot k_i)\,{\dot G}_{ij} -
\epsilon_i\cdot \epsilon_j\,{\ddot G}_{ij} \right]} \times
({\rm anti-hol})\right|_{\rm multi-linear} \ .
\label{kinematicfactor}
\fqr
In \rf{kinematicfactor} 
the dotted $G$s represent worldline derivatives of the complete propagator, 
including bosonic zero modes.\footnote{
This notation follows that of the first-quantized 
form of scattering amplitudes at one-loop, 
derived and motivated by string theory considerations.}  
In \rf{kinematicfactor} we  also have a multiplicative 
factor of the mixing from holomorphic/anti-holomorphic, 
\bqr
\exp{\left[-(\eps_i\cdot{\bar\eps}_j+{\bar\eps}_i\cdot\eps_j)\,H_{ij}\right]}
\ ,
\fqr
where $H_{ij}:=\partial_i {\bar\partial}_j G_{ij}$ in the 
field-theory expression for the amplitude.  This comes from 
the last term in \rf{ampepsilon} and is zero for the $N{=}2$ string because 
of the MHV-type condition that $\eps_i{\cdot}{\bar\eps}_j = 0$ 
for all external legs $i,j$.  The propagator in \rf{kinematicfactor} is 
${\dot G}_{ij} = -{1\over 2}\,{\rm sign}(x_{ij}) + x_{ij}$, and the usual
Feynman parameters are related to $x_i$ via $x_i=\sum_{j=1}^i a_j$.
The point of the form as written in \rf{multilinear} is that the
kinematical expression arises from the standard first-quantized form
of a particle, as generated from integrating over the worldline with
measure factor \rf{taufactor}.  Mapping the $N{=}2$ zero-slope limit to
this expression removes the need to explicitly integrate over
$\tau_2$ (including four-dimensional box integrals with up to eight
insertions of loop momenta in the numerator) because the amplitudes
are known \cite{Bern:1998xc}.  The Feynman-parametrized form of
\rf{multilinear} and \rf{kinematicfactor} produces the integral
form of the zero-slope limit of the $N{=}2$ string expression but with
$d{=}2$ as opposed to $d{=}4$, as we shall now demonstrate.

We begin by writing all possible $\phi^3$ diagrams, obtained by
pinching together different sets of vertex operators.  Then, after 
expanding the kinematical factor in \rf{kinematicfactor} we collect sets of 
${\dot G}_{ij}$ (and ${\dot{\bar G}}_{ij}$) terms in accord with the tree 
and loop rules (example diagrams are illustrated in Figure 3).  
The Bern-Kosower tree rule \cite{Bern:1988tw} in the low-energy extraction 
involves substituting on an external leg $-1/(P_i+P_j)^2$ for the 
occurence of a single power of ${\dot G}_{ij}$, from the outside of the diagram 
into the diagram, and then resubstituting $i=j$ in the remaining momentum flow
of the tree-line as well as making the substitution in the remaining 
${\dot G}_{ij}$ (and ${\dot{\bar G}}_{ij}$) factors.  
In the gravity analog we substitute the same when there is a single product 
of ${\dot G}_{ij} {\dot{\bar G}}_{ij}$.  
In the string-theory amplitude, this amounts to pinching a pole from the 
$e^{-k_i\cdot k_j G_{ij}}$ kinematical factor 
(i.e. integrating near $z_i\sim z_j$).  
After substituting the tree rules on an individual diagram we have a 
remaining kinematical factor on which we apply the loop rules.  
The tree rules do not depend on the spacetime spin of the particle being 
integrated out; rather, the loop rules map to the internal spacetime 
statistics of the particle.  

\begin{figure}[htb]
\vskip-3cm 
\begin{center} 
\epsfig{file=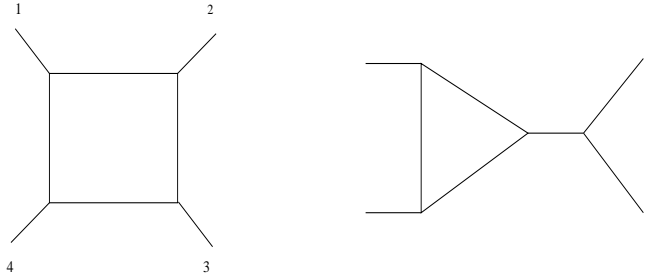,height=10cm,width=9cm,angle=-90} 
\end{center}
\vskip-3cm
\caption{\sl
Example pinched contributions from the string amplitude 
in the field-theory limit.}
\vskip.2cm
\end{figure}

The Bern-Kosower loop rules \cite{Bern:1988tw} tell us to expand 
the holomorphic and anti-holomorphic terms in \rf{kinematicfactor} for a 
given ordering of the four external lines and, after applying the tree 
rules, to substitute the factors of ${\dot G}_{ij}$ by the low-energy 
expansion of the propagators with an overall factor of two for combinatorics, 
\bqr
{\dot G}_{ij}\ \rightarrow\
-{1\over 2}~ {\rm sign}(x_{ij}) + x_{ij} \ . 
\label{ruleone}
\fqr
These generate the uncyclic contributions and represent the bosonic portion 
of the worldsheet correlators in the zero-slope limit.  Next, we examine 
the integrand for cyclic occurances of ${\dot G}_{ij}$, following the ordering 
of the legs exiting the loop 
(for example ${\dot G}_{12} {\dot G}_{23} {\dot G}_{31}$), 
and substitute as follows,
\bqr
{\dot G}_{ij} {\dot G}_{ji}\ \rightarrow\ 2
\qquad{\rm and}\qquad
{\dot G}_{i_1 i_2} {\dot G}_{i_2 i_3} \ldots {\dot G}_{i_n i_1}\ 
\rightarrow 1 \qquad (n>2) \ .
\label{ruletwo}
\fqr
After having applied the cyclic substitutions in \rf{ruletwo} some 
${\dot G}_{ij}$ may be left unsubstituted; they are to be replaced 
according to \rf{ruleone}.  The outcome is the cyclic contributions and 
model the zero-slope limit of the fermionic correlations.  The loop rules 
are applied separately on the ${\dot G}_{ij}$s and ${\dot{\bar G}}_{ij}$s 
in the case of gravity.  

In the case of Einstein-Hilbert gravity, the rules in \rf{ruleone} and 
\rf{ruletwo} generate a graviton as the state within the loop.
In the field-theory limit, the cyclic and uncyclic substitutions arise 
from the fermionic and bosonic worldsheet propagators, respectively 
(when there are no ${\ddot G}_{ij}$ terms), and match with the form of 
${\rm K}_{KN}$.  (A worldline systematics in the case of spin $[J\leq 1]$ 
has also been analyzed in a number of works, including \cite{Strassler:1992zr} 
within the context of 1PI diagrams.) 

There are further simplifications in the integrand of the MHV amplitudes that 
are beyond the naive collecting of terms obtained from expanding the 
Koba-Nielsen factor.  The fact that integrating out a spacetime graviton
or a spacetime complex scalar in the loop makes no difference for the amplitude
implies cancellations of the cyclic terms obtained from the second rule in
\rf{ruletwo}. In order to obtain the MHV amplitude with an internal complex
scalar, only the first loop rule \rf{ruleone} needs to be implemented on both
the holomorphic and anti-holomorphic sides.

In the MHV amplitudes, which satisfy \rf{spectralflow}, this means 
that integrating out the cyclic contributions in \rf{ruletwo} gives identically 
zero ($A^{[2]} = A^{[0]}$).  This fact has been noted in \cite{Bern:1993wt} in 
the application to a gravitational four-point amplitude with helicity 
assignment $(--,++,++,++)$.  At the level of the superstring and the $N{=}2$ 
string this means that the worldsheet fermions do not contribute to the 
MHV amplitudes in the field-theory limit.  Momentum conservation eliminates 
their total sum; this is demonstrated in the next section.  We shall find an 
identical result when integrating over the spin structures prior to taking 
$\alpha'\to0$.

The $q$-expansion of the derivative of the bosonic propagator involved in
extracting $K_{\rm KN}$ is (see the Appendix)
\bqr
\partial \Bigl[ \ln |E(z)|^2 - {2\pi\over \tau_2} ({\rm Im}\,z)^2 \Bigr]
\ =\ i\pi - {2\pi i\over 1- e^{i\alpha} \vert q\vert^u} + {2i\pi\over \tau_2}
 {\rm Im}\,z + {\cal O}(q) \ ,
\label{slopeE}
\fqr
where ${\rm Im}\,z\geq 0$.   Effectively, after the angular integration 
\cite{D'Hoker:1995yr, Chalmers:1998dc}, 
the middle term in \rf{slopeE} integrates 
to $-2\pi i$ with higher-order in $\alpha'$ effects (naively there appears 
to be a potential singularity), yielding the outcome
\bqr
\partial \Bigl[ \ln |E(z)|^2 - {2\pi\over \tau_2} ({\rm Im}\,z)^2 \Bigr]
\ \to\ -i\pi+ {2i\pi\over \tau_2} {\rm Im}\,z \ =\ -i\pi ( 1 - 2y) \ ,
\fqr
in agreement with the rules in \rf{ruleone} and \rf{ruletwo} and the
tensor algebra of the one-loop diagram after angular integration.\footnote{
In the field theory limit $\cot{2\pi z}\rightarrow i$.}

Turning to the fermions, one observes that the field-theory limit of a
Szeg\"o kernel is independent of the spin structure $\ab$. 
In the $q$-expansion,
\bqr
S\ab(z_{i_1 i_2},\tau)\,S\ab(z_{i_2 i_3},\tau) \ldots S\ab(z_{i_n i_1},\tau)
\ \to\  (i\pi)^n + O(q)
\fqr
after the integration over the angular coordinates $\alpha_i$.  With this
substitution, the second rule in \rf{ruletwo} obtains for the
integral expression of the amplitude.  The zero-slope limit of the
$N{=}2$ string reproduces individually {\it all\/} the diagrams of the
gravity amplitude after a careful tracking of the indices of the
${\dot G}_{ij}$ which come in a specific order within the Koba-Nielsen
form in \rf{kinematicfactor} (${\dot G}_{ij} = -{\dot G}_{ji}$).

The primary difference between the integrands of the $N{=}2$ string
and the IIB superstring truncated to obtain four-dimensional gravity 
lays in the integration measure. Concretely,
the $N{=}2$ string scattering amplitude at $n$-point has an
extra factor of $\tau_2$ compared to the amplitude obtained from a
field-theory calculation using the Feynman rules of the self-dual 
gauge theory~\cite{Marcus}. 
As a result, the amplitude in \rf{gravfourrelation} is
obtained effectively in $d{=}2$ and not in $d{=}4$.  

We compare now with the IIB superstring measure, continued to $D$ dimensions. 
On a torus with spin structure $\ab$, the NSR fermions and the supersymmetry
ghosts $(\beta,\gamma)$ produce the determinantal factors 
\bqr
Z_\psi\ab\ =\ 
\vert \vartheta\ab(0,\tau)\vert^D\, \vert \eta(\tau) \vert^{-D}
\qquad\qquad
Z_{\beta\gamma}\ab\ =\ 
\vert \vartheta\ab(0,\tau)\vert^{-2}\, \vert \eta(\tau) \vert^2
\fqr
while the bosonic coordinates and the reparametrization ghosts $(b,c)$ yield
\bqr
Z_x\ =\ \tau_2^{-D/2} \vert\eta(\tau)\vert^{-2D} \qquad\qquad
Z_{bc}\ =\ \tau_2\,\vert \eta(\tau)\vert^4 \ .
\fqr
Together with the Weyl-Peterson measure $d^2\tau/\tau_2^2$,
the product $Z_\psi Z_{\beta\gamma} Z_x Z_{bc}$ yields
\bqr
{d^2\tau\over\tau_2^2} ~\tau_2^{-(D-2)/2} ~
\vert \vartheta\ab(0,\tau)\vert^{D-2}~ \vert \eta(\tau) \vert^{-3(D-2)} \ , 
\label{IIBmeasure}
\fqr 
not taking into account the factors associated with the vertex operators.  
Upon compactification on $T^{D-d}$, it is modified by a lattice sum,  
\bqr  
Z(\Gamma)\ =\ 
 \tau_2^{(D-d)/2} ~\sum_{(P_L,P_R)\in \Gamma} 
e^{i\pi \tau P_L\cdot P_L - i\pi {\bar \tau} P_R\cdot P_R}  
\label{lattice}
\fqr  
where $(P_L,P_R)$ parametrize the $(p,q)$ signature lattice of dimension $D-d$ 
(consistency requires $P_L^2{-}P_R^2\in2\Z$ and $p{-}q\in8\Z$). 
Furthermore, the individual vertex operators generate powers of $\tau_2$ 
after evaluating $d^2z$ (the volume $\int\!d^2z = \tau_2$).  

The two measures, \rf{IIBmeasure} times \rf{lattice} for the compactified 
IIB superstring on one side and \rf{partitionfunction} times \rf{spinmeasure}
for the $N{=}2$ string on the other, differ at zero-slope by a single factor 
of $\tau_2$~\cite{Marcus}:
\bqr
{d^2\tau\over\tau_2^2} ~\tau_2^{-(d-2)/2} 
\qquad\longleftrightarrow\qquad
{d^2\tau\over\tau_2^2} ~\tau_2^{-(d-4)/2} \ .
\fqr
This calculation indicates the dimensional 
shift interpretation of the field-theory integration: the IIB superstring 
compactified on $T^6$ involves a $d^2\tau/\tau_2^{3-n}$ at $n$-point (after 
inserting $D{=}10$ and $d{=}4$ in \rf{IIBmeasure} and \rf{lattice}).  This 
is the same factor that the bosonic string in $d{=}26$ compactified on 
$T^{22}$ generates. 
In contrast, the $N{=}2$ string requires a $d^2\tau/\tau_2^{2-n}$.

%%%%%%%%%%%%%%%%%%%%%%%%%%%%%%%%%%%%%%%%%%%%%%%%%%%%%%%%%%%%%%%%%%%%%%%%%%%%

\section{Spin Structure Summation} 

\noindent
{\bf 6.1\ \ Torus Integrals of Elliptic Functions}

\smallskip

This section pushes the expression for the full $N{=}2$ string scattering
amplitude a step further and also provides an alternative calculation
of its field-theory limit.  Concretely, we explicitly evaluate the integrals 
over the monodromies of the worldsheet fermions, before taking the field 
theory limit. At four-point order this involves integrating over 
spin structures various products of up to four holomorphic Szeg\"o kernels 
(those in \rf{termone} and \rf{termtwo}) 
together with the anti-holomorphic side with the measure in \rf{spinmeasure}.  

For a complex structure $\tau$ of the torus, we first define
(suppressing $\tau$ dependence)
\bqr  
h_n(\{z_{ij}\};u) \ :=\ S\ab(z_{12}) \, S\ab(z_{23}) \cdots S\ab(z_{n1}) \ , 
\label{hndefs}
\fqr 
with 
\bqr  
u\ =\ (\sfrac12{-}\alpha )\, \tau + (\sfrac12{-}\beta) 
\fqr 
denoting the spin-structure dependent zero locus of the Szeg\"o kernel.
By inspecting the zeros and poles of \rf{hndefs} we learn how to
rewrite this expression in terms of prime forms,
\bqr  
h_n\ =\ {E(z_{12}-u)\,E(z_{23}-u) \ldots E(z_{n1}-u) \over 
 [E(-u)]^n\,E(z_{12})\,E(z_{23}) \ldots E(z_{n1})} \ ,
\fqr  
which exposes the single $n$th-order pole in~$u$ at the origin.
The simplest case, $n{=}2$, yields  
\bqr  
h_2(z_{12})\ =\ 
{E(z_{12}-u)^2 \over E(u)^2\,E(z_{12})^2}\ =\
-\wp(z_{12})+\wp(u)\ =\ 
\partial^2 \ln E(z_{12}) - \partial^2 \ln E(u) 
\fqr 
where $\wp(z)$ is the Weierstra\ss\ elliptic function.  Furthermore in the 
coincidence limit $z_{n1}\to0$ one observes that $h_n\to h_{n-1}/z_{n1}$.
Since the spin structure has been encoded in an additional torus variable~$u$,
we have to integrate over~$u$ (with correct measure) the functions $h_n$ 
times their anti-holomorphic relatives. 

With $h_0:=1$ we define the integrals 
\bqr  
f_{n,{\bar n}}(\{z_{ij},\bar{z}_{ij}\})\ :=\ 
\langle\,h_n(\{z_{ij}\};u)\ \bar h_{\bar n}(\{\bar{z}_{kl}\};\bar{u})\,\rangle  
\label{uintegral} 
\fqr 
with measure 
\bqr 
\langle\, \ldots \,\rangle \ :=\  
 \int { du\wedge d{\bar u} \over -2\,i\,\tau_2 }\, \ldots 
\fqr 
which normalizes $\langle1\rangle=1$.  Explicit integration 
for \rf{uintegral} is made possible by the following theorem~\cite{FarkasKra}.  
As $h_n(u) du$ and ${\bar h_{\bar n}(\bar u)} d{\bar u}$ 
are both closed one-forms with zero residue at $u{=}0$ 
we can express the surface integral in terms of period integrals 
over the $a$ and $b$ cycle, 
\bqr  
f_{n,\bar n}\ =\ {i\over 2\tau_2} \Bigl[ 
\oint_a h_n(u) du \oint_b \bar h_{\bar n}(\bar u) d\bar{u} - 
\oint_b h_n(u) du \oint_a \bar h_{\bar n}(\bar u) d\bar{u} 
\Bigr] \ . 
\label{spinintegrals}
\fqr 
We next evaluate the period integrals. 

It is a fact \cite{WW} that an elliptic function with a single $n$th-order 
pole can be expressed as a linear combination of $\wp$ and its derivatives
plus a constant. Hence, expanding $h_n(u)$ around the pole (no residue!) 
we obtain
\bqr
h_n(u) &=& h_n^{(n)} u^{-n} + \ldots + 
h_n^{(3)} u^{-3} + h_n^{(2)} u^{-2} + h_n^{(0)} + {\cal O}(u) 
\non
&=& {\textstyle{(-)^n\over(n{-}1)!}}h_n^{(n)}\wp^{(n-2)}(u) + \ldots 
-\sfrac12 h_n^{(3)}\wp'(u) + h_n^{(2)}\wp(u) + H_n^{(0)}
\label{Pexpansion}
\fqr
with Laurent coefficients $h_n^{(k)}(\{z_{ij}\})$,
where we used $\wp(u)=u^{-2}+O(u^2)$ and
\bqr
{\textstyle{(-)^k\over(k{-}1)!}}\, \wp^{(k-2)}(u)\ =\ 
u^{-k} + G_k\,\del_{k\,\rm even} + O(u)
\label{pder}
\fqr
for $k{\ge}3$.
The holomorphic Eisenstein series
\bqr
G_k\ = \sum_{(m,n)\neq (0,0)} {1\over (m\tau+n)^k}\ =\ 2\,\zeta(k) +
O(e^{2\pi i\tau}) 
\label{holeis}
\fqr
occuring in \rf{pder} for even~$k$ lead to a shift of the constant term
in \rf{Pexpansion},
\bqr
h_n^{(0)}\ \to\ H_n^{(0)}\ =\ 
h_n^{(0)} - G_4\,h_n^{(4)} - G_6\,h_n^{(6)} -\ldots - 
G_{2[{n\over 2}]}\,h_n^{(2[{n\over 2}])} \ .
\fqr

The virtue of the expression \rf{Pexpansion} is that the evaluation
of its period integrals has become almost trivial.
Indeed, since for $k{\ge}3$ the antiderivative of $\wp^{(k-2)}(u)$ is 
$\wp^{(k-3)}(u)$, a doubly-periodic function, 
the integral of $\wp^{(k-2)}(u)$ over a closed loop vanishes.
This observation eliminates all period integrals
except for the last two terms in \rf{Pexpansion}.
Hence, we only require the integrals 
\bqr
\oint_a du\ =\ 1 \qquad\qquad \oint_b du\ =\ \tau 
\fqr
as well as
\bqr 
\oint_a du ~\wp(u)\ =\ -2\eta_1 =\ -G_2  \qquad\qquad
\oint_b du ~\wp(u)\ =\ -2\eta_\tau\ =\ 2\pi i - G_2 \tau
\fqr 
together with their complex conjugates, 
where we have introduced the ``almost-modular'' form (the regulated 
form of the divergent sum in \rf{holeis} for $k=2$) 
\bqr
G_2(\tau)\ =\ 
4 - \sum_{(m,n)\neq (0,0)} {1\over (m\tau{+}n)^2 (2m\tau{+}2n{-}1)}\ .
\fqr
Via \rf{uintegral} and \rf{spinintegrals} this leaves us with only three 
basic non-vanishing spin structure averages,  
\bqr  
\langle 1\rangle\ =\ 1\ , \qquad 
\langle \wp\rangle\ =\ -G_2 +{\pi\over \tau_2}\ , \qquad 
\langle \wp{\bar\wp}\rangle\ =\ G_2{\bar G}_2-{\pi\over\tau_2}(G_2+{\bar G}_2)
\ .   
\fqr 

With these averages we can now compute the integrals \rf{uintegral} as
\bqr
f_{n,{\bar n}}\ &=&\ H_n^{(0)}\,\bar{H}_{\bar n}^{(0)} +
H_n^{(0)}\,\bar{h}_{\bar n}^{(2)}\,\langle\bar\wp\rangle +
\bar{H}_{\bar n}^{(0)}\,h_n^{(2)}\,\langle\wp\rangle +
h_n^{(2)}\,\bar{h}_{\bar n}^{(2)}\,\langle\wp\bar\wp\rangle
\\[1ex] \nonumber
&=&\ \Bigl[ H_n^{(0)} + h_n^{(2)} (-G_2 {+} \pit) \Bigr]\,
\Bigl[ \bar{H}_{\bar n}^{(0)} + 
       \bar{h}_{\bar n}^{(2)} (-\bar{G}_2 {+} \pit) \Bigr]
- h_n^{(2)}\,\bar{h}_{\bar n}^{(2)}\,(\pit)^2 
\fqr
and observe that they do not split chirally. 
In the evaluation of the four-point function we require only the cases 
of $(n,{\bar n})\in\{(0,0),(2,0),(2,2),(4,0),(4,2),(4,4)\}$ together with 
the transposes.  

It remains to list the coefficients $h_n^{(k)}$.  
In general $h_n^{(1)}=0$ and $h_n^{(n)}=(-)^n$.
Here, we only need $h_2^{(k)}$ and $h_4^{(k)}$ for even $k$,
\bqr
H_2^{(0)} &=& h_2^{(0)}\ 
=\ -\wp(z_{12}) = \partial^2 \ln E(z_{12}) + G_2 \ ,
\\[1ex]
h_4^{(2)} &=& \sfrac12 T_4^{-1} {\tilde\partial}^2 T_4 + 2G_2 \ ,
\\[1ex]
H_4^{(0)} &=& h_4^{(0)}-G_4 \ =\ 
\textstyle{1\over24} T_4^{-1} {\tilde\partial}^4 T_4 + 
      G_2\,T_4^{-1} {\tilde\partial}^2 T_4 + 2G_2^2 \ ,
\fqr
The shorthand notation involving $T_4$ 
(generalizable to higher $n$ in this form) is
\bqr
T_4 &=& E_{12}\, E_{23}\, E_{34}\, E_{41} \ ,
\\[2ex]
{\tilde\partial}^2 T_4 &=& 
E_{12}'' E_{23} E_{34} E_{41} + E_{12} E_{23}'' E_{34} E_{41} +
E_{12} E_{23} E_{34}'' E_{41} + E_{12} E_{23} E_{34} E_{41}''
\non
&&+\ 2 E_{12}' E_{23}' E_{34} E_{41} + 2 E_{12}' E_{23} E_{34}' E_{41} +
    2 E_{12}' E_{23} E_{34} E_{41}'
\non
&&+\ 2 E_{12} E_{23}' E_{34}' E_{41} + 2 E_{12} E_{23}' E_{34} E_{41}' +
    2 E_{12} E_{23} E_{34}' E_{41}'
\label{t4}
\fqr
and similarly for ${\tilde\partial}^4 T_4$, where we abbreviated 
$E_{ij}=E(z_{ij})$.  The higher ${\tilde\partial}^k$ represents the 
actions of $k$ derivatives spread out with respect to the insertion 
points $z_{ij}$.  

For later reference, we present the first few spin structure integrals:
\bqr
f_{2,0} &=& \partial^2 \ln E_{12} + \pit
\label{f20}
\\[1ex]
f_{2,2} &=& \vert\,\partial^2 \ln E_{12} + \pit\,\vert^2 - (\pit)^2
\label{f22}
\\[1ex]
f_{4,0} &=& \textstyle{1\over24} T_4^{-1} {\tilde\partial}^4 T_4 +
( G_2 {+} \pit)\,\sfrac12 T_4^{-1} {\tilde\partial}^2 T_4 + 2G_2\,\pit
\ .
\label{f40}
\fqr
The analysis above is generalizable to higher genus by employing the 
prime forms pertaining to the higher-genus Riemann surface.

%%%%%%%%%%%%%%%%%%%%%%%%%%%%%
\bigskip

\noindent
{\bf 6.2\ \ Zero-slope Limit}
\smallskip

We now analyze the field theory limit of the various terms obtained 
from summing over the spin structures. In the process of evaluating the ratios 
of prime forms $E(z_{ij})$ and their derivatives, the following can be 
implemented: 
\bqr  
E(z)\ \rightarrow\ {1\over 2\pi} \sin(2\pi z) \ , 
\label{Elimit} 
\fqr 
which in $f_{n,{\bar n}}$ leads to products of unity and $\cot(2\pi z)$.  
After the angular integration over $z_r$, $\cot(2\pi z)\rightarrow i$.  
Therefore, the analysis of the fermionic correlator terms $f_{n,{\bar n}}$ 
reduces to combinatoric factors and derivatives of \rf{Elimit} with respect 
to the $z$-coordinates, together with the appropriate $\phi^3$ diagram 
via pinching the $\prod \vert E(z_{ij})\vert^{-\alpha's_{ij}}$.   

We consider first $f_{2,0}(z)$ and $f_{2,2}(z,{\bar z})$ 
given by \rf{f20} and \rf{f22}, respectively,  
\bqr  
f_{2,0}(z)\ \rightarrow\ 
-(2\pi)^2 \bigl[ 1+ \cot^2(2\pi z)\bigr] + {\pi\over\tau_2}
\ \rightarrow\  {\pi\over\tau_2} \ ,  
\label{20function}
\fqr 
\bqr 
f_{2,2}(z,\bar z)\ \rightarrow\ 
\Bigl| -(2\pi)^2 \cot^2(2\pi z) - (2\pi)^2 + {\pi\over\tau_2}\, \Bigr|^2
- \Bigl({\pi\over\tau_2}\Bigr)^2  
\ \rightarrow\ 0 \ , 
\label{22function} 
\fqr 
where $\bar z$ is a variable independent of $z$.
All of the contributions containing $f_{2,0}$ vanish after adding them up in 
the expansion of the Koba-Nielsen factor; 
we will show this after analyzing the remaining terms. 

The remaining integrals $f_{n,{\bar n}}$ all involve at least either $n=4$ or 
${\bar n}=4$.  The field theory limit of such a term, exemplified in \rf{f40}, 
does not vanish individually in the kinematical expression, 
but like terms add up to zero as we will show below.  
In the $\tau_2\rightarrow\infty$ limit the Eisenstein series simplify,
\bqr 
G_2\ \to\ {\pi^2\over 3} \qquad{\rm and}\qquad G_4\ \to\ {\pi^4\over 45} \ . 
\fqr 
As displayed in \rf{t4}
the ${\tilde\partial}^4$ and ${\tilde\partial}^2$ derivatives produce a large 
number of products of $E^{(k)}_{ij}/E_{ij}$.  However, in 
the field theory limit no $z_{ij}$ dependence survives and
\bqr
{E^{(k)}_{ij} \over E_{ij}}\ \to\ (2\,\pi\,i)^k  
\fqr
which yields
\bqr
T_4^{-1} {\tilde\partial}^4 T_4 \ \rightarrow\ +256\,(2\pi)^4 \qquad\qquad
T_4^{-1} {\tilde\partial}^2 T_4 \ \rightarrow\  -16\,(2\pi)^2 \ .
\fqr  

Collecting all terms, the net limits of the remaining terms are 
\bqr  
f_{4,0}\ &\rightarrow&\ {\tilde f}_{4,0}\ =\ 
10\,(2\pi)^4 - {47\over6} (2\pi)^2 {\pi\over\tau_2} 
\label{40function}
\\[1ex]
f_{4,2}\ &\rightarrow&\ {\tilde f}_{4,2}\ =\
10\,(2\pi)^4 {\pi\over\tau_2}
\label{42function}
\\[1ex]
f_{4,4}\ &\rightarrow&\ {\tilde f}_{4,4}\ =\
100\,(2\pi)^8 - {470\over3} (2\pi)^6 {\pi\over\tau_2} \ . 
\label{44function}
\fqr 
It is interesting that a $1/\tau_2$ appears in these terms, which indicates 
the modification necessary to obtain the MHV amplitudes.  These terms  
terms do not produce a $0/0$ effect as they also vanish in four dimensions, 
being proportional to the difference between a scalar contribution and a 
graviton contribution to the MHV amplitude.  

The spin structure integrals $f_{n,\bar n}$ are being multiplied by 
kinematical factors $t_{n,\bar n}(\{\eps_i,k_j\})$ stemming from the 
contractions of polarization and momentum vectors on four vertex 
operators~\rf{normalizedvertex}.  Since in the field-theory limit all 
$z$ dependence has dropped from ${\tilde f}_{n,\bar n}$, the latter can 
be factored out from the remaining integrations.  This fact allows one 
to combine directly the various permutations of a given kinematical factor.

We now analyze the kinematical factors $t_{n,\bar n}=t_n\bar t_{\bar n}$.
We begin with the contractions of four pairs of fermions, which can happen 
in two distinct ways.  The first possibility is a single cycle connecting 
all pairs,
\bqr 
t_4^{(1234)}\ =\ 
\eps_1\cdot k_2\, \eps_2\cdot k_3\, \eps_3\cdot k_4\, \eps_4\cdot k_1 \ ,
\fqr 
together with permutations $(1\leftrightarrow 2)$ and $(2\leftrightarrow 3)$.  
With arbitrary reference momenta $q$ chosen the same for all polarization 
vectors, we have  
\bqr  
t_4^{(1234)}\ =\ [12] [23] [34] [41] 
\fqr 
which follows from the substitution 
\bqr  
\eps^{\alpha\dot\alpha}(k,q)\ =\ 
i {q^\alpha k^{\dot\alpha}\over q^\beta k_\beta} \ . 
\label{polarization}
\fqr 
Adding the three permutations produces
\bqr  
t_4^{(1234)}+t_4^{(2134)}+t_4^{(1324)} =\ 
[12] [23] [34] [41] + [21] [13] [34] [42] + [13] [32] [24] [41] \,.
\label{cyclicsum}
\fqr 
Employing twice the Fierz identity
\bqr
[AB] [CD]\ =\ [AC] [BD] + [AD] [CB]
\label{fierz}
\fqr
we find
\goodbreak
\bqr
t_4^{(1234)}+t_4^{(2134)}+t_4^{(1324)}  &\propto& 
[12] [23] [34] [41] + [13] [24] \Bigl( [12] [34] + [14] [23] \Bigr)
\non &=& 
[12] [23] [34] [41] + [13]^2 [24]^2
\non &=&
[12] [34] \Bigl( [24] [31] + [21] [43] \Bigr) + [13]^2 [42]^2
\non &=&
- [12] [24] [43] [31] + [12]^2 [34]^2 + [13]^2 [42]^2 \ .
\label{twofierz}
\fqr
Symmetrizing both sides of this equation and using the identity
\bqr
[13]^2 [42]^2 + [12]^2 [43]^2 + [14]^2 [32]^2\ =\ 0
\label{ident}
\fqr
(again from \rf{fierz}) one discovers that 
$t_4^{(1234)}+t_4^{(2134)}+t_4^{(1324)}$ equals minus itself.
Thus, the sum in \rf{cyclicsum} vanishes.

The second option for contracting four pairs of fermions produces
two cycles of two pairs each,
\bqr  
t_4^{(12)(34)}\ =\
\eps_1\cdot k_2\,\eps_2\cdot k_1\,\eps_3\cdot k_4\,\eps_4\cdot k_3 \ ,
\fqr 
together with its two permutations.  Via \rf{polarization} this equals 
\bqr  
t_4^{(12)(34)}\ =\ [12]^2 [34]^2 \ , 
\fqr 
which upon adding the three permutations and using \rf{ident} equals zero, too. 
Similar additions of the field theory limit for the fermionic terms add to 
zero in the three-point and two-point $\phi^3$ diagrams, where the momentum 
structure involves three and two independent momenta, respectively.    

The previous analysis regarding the cyclic terms proportional to $f_{4,0}$ 
generalizes in a straightforward manner to the remaining cyclic terms 
multiplying $f_{4,2}$ and $f_{4,4}$.  In the field-theory limit all of the 
$z$ dependence on the holomorphic half of the kinematical expression 
multiplying these functions is absent (the anti-holomorphic half multiplying 
$f_{4,2}$ includes bosonic zero modes which translate to Feynman 
parameters in the field theory limit).  After summing over the different 
contributions on the holomorphic half, these contributions equal zero, 
as was shown in the preceeding paragraphs.  

Finally we analyze the $f_{2,0}$ (and $f_{2,2}$) terms.  
The Wick contractions of two pairs of fermions yield a kinematical factor of
\bqr
t_2^{(ij)}\ =\ \eps_i\cdot k_j\, \eps_j\cdot k_i \ ,
\fqr
which multiplies the remaining kinematical structure 
from $\partial_k G_{k\ell}$, as displayed by the solid lines in Figure 2.  
The relation $A^{[0]}=A^{[1]}=A^{[2]}$ enforces all these terms to be zero.  
This ``supersymmetry identity'' 
implies that the fermionic contractions associated with rule two all
generate zero, as discussed in the previous section.  In the string amplitude 
this means that the holomorphic sum of all of the world-sheet fermionic 
correlators with equal weight add to zero.  We have already showed by momentum 
conservation (for the particular MHV helicity structure) that the four-fermion 
terms are zero, it follows that the $t_2^{(ij)}$ terms also add to zero 
(as they all have the same coefficient in \rf{20function}).  This cancellation 
occurs separately for the four-fermion-pair contractions, i.e. the $t^4$ terms, 
as well as the two-fermion-pair contractions, in both two and four dimensions 
(as the supersymmetry idendity holds in both cases).  
The factors of $\tau_2$ in \rf{20function} and \rf{40function}--\rf{44function} 
cause a dimensional shift in the integration.  

%%%%%%%%%%%%%%%%%%%%%%%%%%%%%%%%%%%%%%%%%%%%%%%%%%%%%%%%%%%%%%%%%%%%%%%%%%%%

\section{Discussion} 

In this work we have analyzed several aspects of the quantum scattering 
of the closed $N{=}(2,2)$ closed superstring at genus one.  
First, we have derived the zero-slope limit of the one-loop four-point function 
in the RNS formulation, and explicitly integrated over the spin structures 
of the worldsheet fermions. We have found agreement with the existing vanishing 
theorems in the literature. The mapping of the genus-one moduli space integrand 
to an MHV amplitude at $n$-point order is performed.  Second, we have compared 
the one-loop integrated three- and higher-point string amplitudes with those of
self-dual gravity. The disagreement (vanishing versus nonzero MHV) could be
traced to a known~\cite{Marcus} difference in the integration measure 
whose origin is the local R symmetry of the $N{=}2$ string.  Third, 
we have made manifest the Lorentz and coordinate invariance of the quantum 
(and classical) scattering by normalizing the vertices and incorporating 
spinor helicity techniques.  Most of this analysis carries over 
straightforwardly to the open string.  

A number of new features have arisen regarding the quantum amplitudes.   
The $N{=}2$ string has field equations of self-duality at the 
classical level, but at genus one its amplitudes are not directly 
found from self-dual field theory in four dimensions.  Rather, the result 
appears in the loop integration as the dimensionally regulated version
of the self-dual amplitudes, continued to two dimensions 
(with external kinematics in two complex dimensions). 
These field-theory amplitudes indeed vanish.  

The two-dimensional nature of the $N{=}2$ string loop integration suggests 
that the effective dynamics of this string is only (real) two-dimensional. 
Then, the vanishing of two-dimensional gravity (and Yang-Mills) amplitudes 
may account for the all-order vanishing of the string amplitudes (like at 
genus one).   
Clearly, a string in two-dimensional target spacetime has 
no room for physical excitations (at generic momenta).  In four-dimensional 
spacetime, however, the same situation can be arrived at by increasing the 
worldvolume dimension from two to four, since a spacetime-filling brane 
affords only topological degrees of freedom.  Indeed, the analogy
\bqr
T\ \longleftrightarrow\ J \qquad\qquad\qquad\qquad
(b,c)\ \longleftrightarrow\ (b',c') \qquad\qquad
\fqr
and the fact that the $N{=}2$ string ghost systems remove two {\it complex\/}
unphysical directions from the excitation spectrum (best seen for the
RNS fermions) have led to the speculation~\cite{Ooguri:1991fp} that the
$N{=}2$ string actually is a space-filling brane. It is tempting to
interpret the $U(1)^2$ fibre associated with the local R symmetry as
carrying the two additional dimensions, making for a total of four
parametrizing the full bundle. The $N{=}2$ string formulation then amounts
to a fibration of the $2{+}2$ dimensional world-volume over a Riemann surface.

Another avenue is to search for modifications in the string amplitude
which resurrect the non-vanishing four-dimensional MHV scattering.
A single factor of $1/\tau_2$ in the integrand of the closed string is 
required to extract the one-loop self-dual field-theory amplitudes in $d{=}4$.  
An insertion of an unintegrated zero-momentum vertex operator or a bosonic 
zero mode would already do the job, for example, through 
\bqr 
\mathop{\rm lim}_{k=0}~\sqrt{g}~\partial x {\bar\partial} x e^{ik\cdot x} 
\qquad\qquad{\rm or}\qquad\qquad 
\partial{\bar\partial} G(z,{\bar z})\ =\ 
{2\pi\over \tau_2}\,\delta^{(2)}(z,\bar z) \ . 
\fqr 
This conformal anomaly may have a target spacetime interpretation as a 
$\beta$ function expansion around $d{=}2$.  The insertion of the zero mode, 
breaking the worldsheet conformal invariance, expands the amplitude to 
those of one-loop self-dual field theory, and through perturbations of 
self-duality to gravity and Yang-Mills theory.  

We have analyzed one-loop string amplitudes in the field-theory limit, 
i.e. calculated the leading term in the $q$-expansion of a string amplitude.  
As the vanishing theorems and the Ward identities of the $N{=}2$ string
imply that the entire tower of $q$-expansion coefficients is zero,
we expect the above two-dimensional interpretation to hold for the full
$N{=}2$ string theory. A direct verification of the vanishing of the higher-$q$ 
components of the genus one string amplitude is outside the scope of the 
present paper; however, we have made important steps in that direction by 
providing the reader with an explicit expression of the string integrand after 
spin-structure summation. Whether the analysis remains feasible at 
$\alp'{\neq}0$ at the level of the full $q$-expansion is to be shown.

Finally, the one-loop amplitudes generated by the closed $N{=}2$ string 
are related through an order $\epsilon{=}10{-}d$ identity to those 
of IIB supergravity in ten dimensions. 
Via a relation $A_{N=2} \sim \epsilon A_{IIB}$, 
the zero-slope limit of the $N{=}2$ string captures the ultraviolet portion
of the IIB amplitudes; the latter amplitudes are finite 
in a dimensionally regulated form in ten dimensions.  As both amplitudes 
are low-energy limits of critical string theories, this suggests a relation 
between the $N{=}2$ and $N{=}1$ strings, which at multi-loop requires a similar 
relation between the MHV amplitudes and the non-MHV IIB amplitudes.  It 
is interesting to note that membrane-string and string-string connections have 
been noted in the context of the heterotic $(2,1)$ formulation in relation 
to world-volumes of membranes \cite{Kutasov:1996fp,Kutasov:1997vh}.

\goodbreak
\vskip .3in 
\noindent {\it Acknowledgements} 
\vskip .15in

The work of G.C. is supported in part by the U.S. Department of 
Energy, Division of High Energy Physics, Contract W-31-109-ENG-38.  
Support for O.L. and B.N. by the German National Science Foundation (DFG)
under grant LE 838/5-2 is gratefully acknowledged.
G.C. thanks the Institute for Theoretical Physics at the University 
of Hannover for hospitality and Warren Siegel for relevant 
correspondence. O.L. acknowledges fruitful discussions with 
Michael Flohr, Klaus Hulek, Jacob Nielsen and Jeroen Spandaw.
B.N. is thankful to Klaus J\"unemann for useful conversations.

%%%%%%%%%%%%%%%%%%%%%%%%%%%%%%%%%%%%%%%%%%%%%%%%%%%%%%%%%%%%%%%%%%%%%%%%%%%

\vskip .5in
\section{ Appendix : Theta Functions}

We list in this Appendix some of the properties of the Jacobi theta functions
and elliptic functions useful in this work.  The theta function with
$(\alpha,\beta)$ characteristics is defined by the infinite sum

\bq
\vartheta\ab (z,\tau)\ =\
\sum_{n\in Z} e^{\pi i \tau (n+\alpha)^2
 + 2\pi i(n+\alpha)(z+\beta)} \ .
\fq
The theta function satisfies the identity
\bqr
\vartheta\ab (z,\tau)
&=& e^{\pi i\tau\alpha^2+2\pi i\alpha(z+\beta)} ~
\vartheta\oo (z+\tau\alpha+\beta, \tau) 
\non
&=& e^{\pi i\tau (\alpha^2-1/2)^2+2\pi i(\alpha-1/2)(z+\beta)} ~
\vartheta\hh (z+(\alpha{-}\sfrac12)\tau+(\beta{-}\sfrac12), \tau) 
\fqr
with the $\vartheta\hh (-z,\tau)=-\vartheta\hh (z,\tau)$.
Abbreviating $q=e^{2\pi i\tau}$,
the infinite product form of the odd theta function reads
\bq
\vartheta\hh (z,\tau)\ =\ i q^{1/8} e^{\pi i z}
\prod_{n=1}^\infty \left(1-q^n\right)
\prod_{n=0}^\infty \left[\left(1-q^ne^{-2\pi i z}\right)
\left(1-q^{n+1}e^{2\pi i z}\right)\right] \ ,
\fq
and that of its $z$-derivative at $z{=}0$ is
\bq
\vartheta^\prime\hh (0,\tau)\ =\ -2\pi q^{1/8}
\prod_{n=1}^\infty \left(1-q^n\right)^3 \ .
\fq

In the same manner, we may rewrite the prime form as
\bq
E(z,\tau)\ =\ {\vartheta\hh (z,\tau) \over \vartheta'\hh (0,\tau)}
\ =\ {e^{\pi i z}\over{2\pi i}}\, 
{\prod_{n=0}^\infty \left[\left(1-q^ne^{-2\pi i z}\right)
\left(1-q^{n+1}e^{2\pi i z}\right)\right] \over
\prod_{n=1}^\infty \left(1-q^n\right)^2} \ ,
\fq
where inspection reveals that
\bqr
E(z{+}1,\tau)\ =\ -E(z,\tau) \qquad{\rm and}\qquad 
E(z{+}\tau,\tau)\ =\ -e^{-\pi i\tau-2\pi iz} E(z,\tau) \ .
\fqr
The chiral bosonic correlator (without the zero-mode part) is
\bqr
G(z,\tau)\ =\ - \ln E(z,\tau)\ =\
- \ln {\vartheta\hh (z,\tau) \over \vartheta^\prime\hh (0,\tau) } \ .
\fqr
Once we insert the bosonic propagators into the expression for the four-point
function, the $z$-independent factors will vanish as a result of momentum
conservation.  We also need the expanded version of $\partial G$, where
we define the parameter $w=e^{2\pi iz}$,
\bqr
\partial\,G(z,\tau) &=& - \pi i {1+w^{-1}\over 1-w^{-1}} +
 2\pi i \sum_{n=1}^\infty q^n \left( {w\over 1-q^nw}-
 {w^{-1}\over 1-q^nw^{-1}} \right)\cr
&=& -\pi \cot (\pi z) +
 2\pi i \sum_{n=1}^\infty q^n \left( {w\over 1-q^nw}-
 {w^{-1}\over 1-q^nw^{-1}} \right) \ .
\fqr

The fermionic Szeg\"o kernel for a given spin structure
$(\alpha,\beta)\neq(\half,\half)$  is
\bqr
S\ab (z,\tau) &=&
 {\vartheta\ab (z,\tau)\ \vartheta^\prime\hh (0,\tau) \over
  \vartheta\ab (0,\tau)\ \vartheta\hh (z,\tau) } 
\non
&=& e^{2\pi i (\alpha-1/2)z} ~
{ \vartheta\hh (z+(\alpha{-}\sfrac12)\tau+(\beta{-}\sfrac12),\tau)\ 
  \vartheta^\prime\hh (0,\tau) \over
  \vartheta\hh ((\alpha{-}\sfrac12)\tau+(\beta{-}\sfrac12),\tau)\ 
  \vartheta\hh (z,\tau) } \ .
\fqr
For the odd spin structure, the fermionic propagator (again ignoring the
zero-mode part) is a derivative of the chiral bosonic Greens function,
\bqr
S\hh (z,\tau)\ =\
-\partial\, G(z,\tau)\ =\ \partial\, \ln E(z,\tau)\ =\ 
{ \vartheta^\prime\hh (z,\tau) \over \vartheta\hh (z,\tau) } \ .
\fqr
This relation between bosonic and fermionic propagator extends to the
anti-holo\-mor\-phic part and the zero-mode part as well, 
\bqr
{2\pi\over i\tau_2}\,{\rm Im}\,z\ =\ 
\partial_z\,{2\pi\over \tau_2} \left[ {\rm Im}\,z \right]^2 \ .
\fqr
We must integrate over all spin structures including the odd one;
however, the odd spin-structure correlators do not contribute to
any of the amplitudes derived in this work.

The Szeg\"o kernels $S\ab(z,\tau)$ as well as $\partial G(z,\tau)$
are singular as we take $z\rightarrow 0$; however, the combination
\bqr
{\cal F}\ab (z,\tau)\ =\
S\ab (z,\tau) - \partial\,G(z,\tau)
\fqr
is finite in this limit.

The Weierstra{\ss} function $\wp(z,\tau)$ is the unique doubly periodic 
function with a single second-order pole at the origin and no constant term
in its Laurent expansion,
\bqr
\wp(z) & = & {1\over z^2}\ +\ 
\sum_{(m,n)\neq(0,0)} \left( {1\over (z{-}m\tau{-}n)^2} - {1\over z^2} \right)
\non
&=& {1\over z^2}\ +\ \sum_{k=1}^\infty\,(2k{+}1)\,G_{2k+2}(\tau)\,z^{2k} \ ,
\fqr
with the modular form
\bqr
G_{2k+2}(\tau)\ =\ \sum_{(m,n)\neq (0,0)} (m\tau{+}n)^{-(2k+2)}
\ =\ 2\,\zeta(2k{+}2) + O(q)
\label{eisenstein}
\fqr
being known as the holomorphic Eisenstein function of weight $2k{+}2$, 
for $k{\ge}1$.  
The antiderivative of the $\wp$ function is denoted a $-\zeta(z)$;
it takes the half-point values
\bqr
\zeta(1/2)\ \equiv\ \eta_1\ =\ \sfrac12 G_2 \qquad{\rm and}\qquad
\zeta(\tau/2)\ \equiv\ \eta_\tau\ =\ \sfrac12 G_2\tau - i\pi
\fqr
where the failure of the sum in \rf{eisenstein} to absolutely converge
for $k{=}0$ necessitates a regularized definition of the
``almost-modular'' form
\bqr
G_2(\tau)\ =\ 
4 - \!\sum_{(m,n)\neq (0,0)} {1\over (m\tau{+}n)^2 (2m\tau{+}2n{-}1)}\
=\ {4\pi\over i}\,\partial_\tau\,\ln\eta(\tau) \ .
\fqr
The last equality makes contact with the logarithm of the 
Dedekind eta function,
\bqr
\ln\eta(\tau)\ =\ 
\sum_{n=1}^\infty \ln\,(1-e^{2\pi in\tau}) + i\pi{\textstyle{\tau\over12}} \ ,
\fqr 
and generalizes to the higher Eisenstein functions, e.g.
\bqr
2\,G_2(\tau)^2 - 10\, G_4(\tau)\ =\ 
\Bigl({4\pi\over i} \Bigr)^2\,\partial^2_\tau\,\ln\eta(\tau) \ .
\fqr

\vfill\eject

%%%%%%%%%%%%%%%%%%%%%%%%%%%%%%%%%%%%%%%%%%%%%%%%%%%%%%%%%%%%%%%%%%%%%%%%%

\end{document}